\RequirePackage[bookmarksnumbered,unicode]{hyperref}
\RequirePackage{hyperxmp}

\documentclass[noxmp,sigconf]{acmart}
\usepackage{multirow}
\usepackage{booktabs}
\usepackage[toc,page]{appendix}
\usepackage{amsmath}
\usepackage{pifont}
\usepackage{diagbox}
\usepackage{balance}
\AtBeginDocument{%
  }

\usepackage{tikz}
\usepackage{pgfplots}
\pgfplotsset{compat=newest}
\usepackage{xcolor}
\usepgfplotslibrary{groupplots}

\pgfplotsset{
    colormap={cold}{
        rgb(0cm)=(0.90,0.95,1.0);
        rgb(1cm)=(0.0,0.0,1.0)
    }
}

\begin{document}

\title{TransAct V2: Lifelong User Action Sequence Modeling on Pinterest Recommendation}



\author{Xue Xia}
\email{xxia@pinterest.com}
\affiliation{%
  \institution{Pinterest}
  \city{San Francisco}
  \state{CA}
  \country{USA}
}

\author{Saurabh Vishwas Joshi}
\email{sjoshi@pinterest.com}
\affiliation{%
  \institution{Pinterest}
  \city{San Francisco}
  \state{CA}
  \country{USA}
}

\author{Kousik Rajesh}
\email{krajesh@pinterest.com}
\affiliation{%
  \institution{Pinterest}
  \city{San Francisco}
  \state{CA}
  \country{USA}
}

\author{Kangnan Li}
\email{kangnanli@pinterest.com}
\affiliation{%
  \institution{Pinterest}
  \city{San Francisco}
  \state{CA}
  \country{USA}
}

\author{Yangyi Lu}
\email{yangyilu@pinterest.com}
\affiliation{%
  \institution{Pinterest}
  \city{San Francisco}
  \state{CA}
  \country{USA}
}

\author{Nikil Pancha}
\authornote{Work done at Pinterest.}
\email{npancha@pinterest.com}
\affiliation{%
  \institution{Pinterest}
  \city{San Francisco}
  \state{CA}
  \country{USA}
}

\author{Dhruvil Deven Badani}
\email{dbadani@pinterest.com}
\affiliation{%
  \institution{Pinterest}
  \city{San Francisco}
  \state{CA}
  \country{USA}
}

\author{Jiajing Xu}
\email{jiajing@pinterest.com}
\affiliation{%
  \institution{Pinterest}
  \city{San Francisco}
  \state{CA}
  \country{USA}
}

\author{Pong Eksombatchai}
\email{pong@pinterest.com}
\affiliation{%
  \institution{Pinterest}
  \city{San Francisco}
  \state{CA}
  \country{USA}
}







\renewcommand{\shortauthors}{Xue Xia et al.}


\newcommand{\td}[1]{{\bf\color{red}[{\sc TODO:} #1]}}
\newcommand{\nikil}[1]{{\bf\color{red}[{\sc Nikil:} #1]}}
\newcommand{\xue}[1]{{\bf\color{red}[{\sc Xue:} #1]}}
\newcommand{\yangyi}[1]{{\bf\color{purple}[{\sc Yangyi:} #1]}}

\newcommand{\mA}{\boldsymbol{A}}
\newcommand{\mB}{\boldsymbol{B}}
\newcommand{\mC}{\boldsymbol{C}}
\newcommand{\mD}{\boldsymbol{D}}
\newcommand{\mE}{\boldsymbol{E}}
\newcommand{\mF}{\boldsymbol{F}}
\newcommand{\mH}{\boldsymbol{H}}
\newcommand{\mI}{\boldsymbol{I}}
\newcommand{\mO}{\boldsymbol{O}}
\newcommand{\mP}{\boldsymbol{P}}
\newcommand{\mR}{\boldsymbol{R}}
\newcommand{\mS}{\boldsymbol{S}}
\newcommand{\mU}{\boldsymbol{U}}
\newcommand{\mV}{\boldsymbol{V}}
\newcommand{\mW}{\boldsymbol{W}}
\newcommand{\mM}{\boldsymbol{M}}
\newcommand{\mX}{\boldsymbol{X}}
\newcommand{\mY}{\boldsymbol{Y}}

\newcommand{\mSigma}{\boldsymbol{\Sigma}}
\newcommand{\mLambda}{\boldsymbol{\Lambda}}
\newcommand{\mPsi}{\boldsymbol{\Psi}}
\newcommand{\mPhi}{\boldsymbol{\Phi}}
\newcommand{\matV}{\boldsymbol{\mathrm{V}}}
\newcommand{\matU}{\boldsymbol{\mathrm{U}}}
\newcommand{\matSigma}{\boldsymbol{\Sigma'}}

\newcommand{\va}{\boldsymbol{a}}
\newcommand{\vA}{\boldsymbol{A}}
\newcommand{\vb}{\boldsymbol{b}}
\newcommand{\vc}{\boldsymbol{c}}
\newcommand{\ve}{\boldsymbol{e}}
\newcommand{\vf}{\boldsymbol{f}}
\newcommand{\vg}{\boldsymbol{g}}
\newcommand{\vh}{\boldsymbol{h}}

\newcommand{\vS}{\boldsymbol{S}}
\newcommand{\vH}{\boldsymbol{H}}
\newcommand{\vF}{\boldsymbol{F}}
\newcommand{\vl}{\boldsymbol{l}}
\newcommand{\vO}{\boldsymbol{O}}
\newcommand{\vo}{\boldsymbol{o}}
\newcommand{\vp}{\boldsymbol{p}}
\newcommand{\vq}{\boldsymbol{q}}
\newcommand{\vr}{\boldsymbol{r}}
\newcommand{\vu}{\boldsymbol{u}}
\newcommand{\vw}{\boldsymbol{w}}
\newcommand{\vx}{\boldsymbol{x}}
\newcommand{\vy}{\boldsymbol{y}}
\newcommand{\vz}{\boldsymbol{z}}
\newcommand{\vze}{\boldsymbol{0}}
\newcommand{\vone}{\boldsymbol{1}}

\newcommand{\vxi}{\boldsymbol{\xi}}
\newcommand{\vmu}{\boldsymbol{\mu}}
\newcommand{\vlambda}{\boldsymbol{\lambda}}
\newcommand{\vpsi}{\boldsymbol{\psi}}
\newcommand{\vphi}{\boldsymbol{\phi}}
\newcommand{\vsigma}{\boldsymbol{\sigma}}
\newcommand{\veps}{\boldsymbol{\varepsilon}}
\renewcommand{\epsilon}{\varepsilon}

\newcommand{\calA}{\mathcal{A}}
\newcommand{\calB}{\mathcal{B}}
\newcommand{\calD}{\mathcal{D}}
\newcommand{\calE}{\mathcal{E}}
\newcommand{\calL}{\mathcal{L}}
\newcommand{\calM}{\mathcal{M}}
\newcommand{\calN}{\mathcal{N}}
\newcommand{\calO}{\mathcal{O}}
\newcommand{\calP}{\mathcal{P}}
\newcommand{\calR}{\mathcal{R}}
\newcommand{\calT}{\mathcal{T}}
\newcommand{\calX}{\mathcal{X}}
\newcommand{\calY}{\mathcal{Y}}
\newcommand{\calZ}{\mathcal{Z}}





    

\newcommand{\tikzxmark}{%
\tikz[scale=0.23] {
    \draw[line width=0.7,line cap=round] (0,0) to [bend left=6] (1,1);
    \draw[line width=0.7,line cap=round] (0.2,0.95) to [bend right=3] (0.8,0.05);
}}
\newcommand{\tikzcmark}{%
\tikz[scale=0.23] {
    \draw[line width=0.7,line cap=round] (0.25,0) to [bend left=10] (1,1);
    \draw[line width=0.8,line cap=round] (0,0.35) to [bend right=1] (0.23,0);
}}

\begin{abstract}

Modeling user action sequences has become a popular focus in industrial recommendation system research, particularly for Click-Through Rate (CTR) prediction tasks. However, industry-scale CTR models often rely on short user sequences, limiting their ability to capture long-term behavior. Additionally, these models typically lack an integrated action-prediction task within a point-wise ranking framework, reducing their predictive power. They also rarely address the infrastructure challenges involved in efficiently serving large-scale sequential models.
In this paper, we introduce TransAct V2, a production model for Pinterest's Homefeed ranking system, featuring three key innovations: (1) leveraging very long user sequences to improve CTR predictions, (2) integrating a Next Action Loss function for enhanced user action forecasting, and (3) employing scalable, low-latency deployment solutions tailored to handle the computational demands of extended user action sequences.
To overcome latency and storage constraints, we leverage efficient data-processing strategies and model-serving optimizations, enabling seamless industrial-scale deployment. Our approach’s effectiveness is further demonstrated through ablation studies.
Furthermore, extensive offline and online A/B experiments confirm major gains in key metrics, including engagement volume and recommendation diversity, showcasing TransAct V2's real-world impact.
\end{abstract}



\maketitle

\section{Introduction}


In the digital age, recommender systems have become indispensable tools for navigating the vast sea of online information. By offering personalized recommendations, these systems not only enhance user experience but also drive engagement and revenue growth for businesses across industries.

\begin{figure}[!ht]
  \vspace{-5pt}
  \centering
  \includegraphics[width=0.35\linewidth]{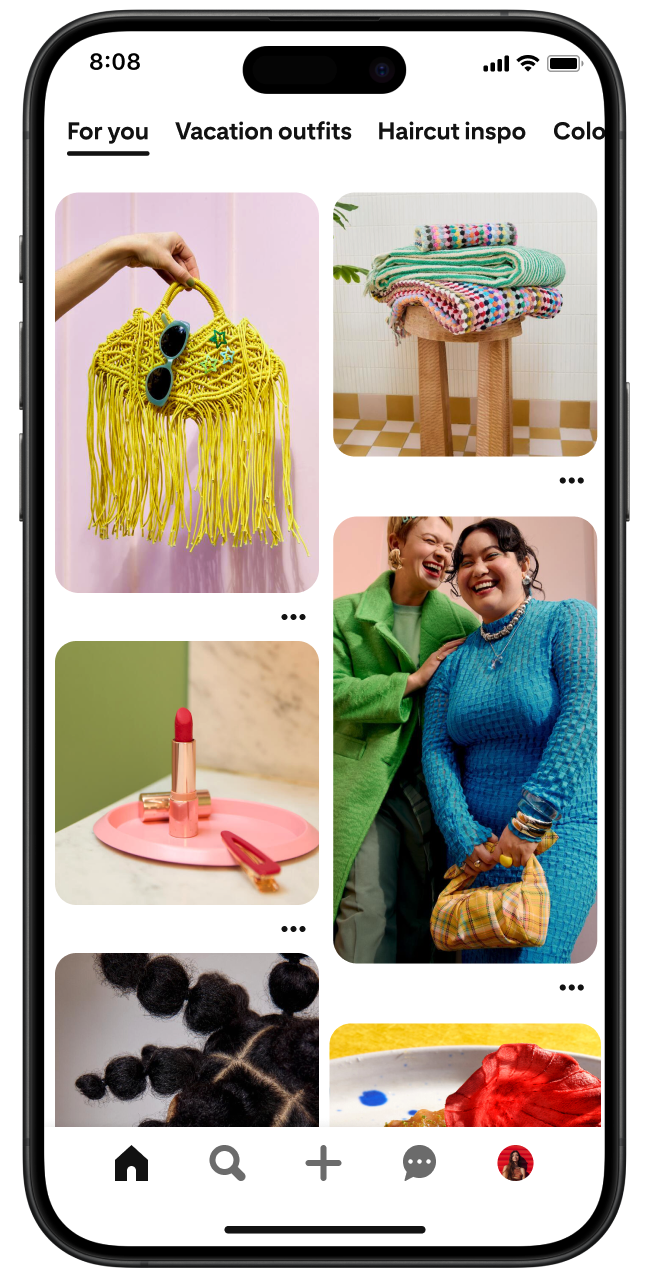}
  \caption{Pinterest Homefeed Page}
  \label{fig:hf}
  \Description{Pinterest Homefeed page}
  \vspace{-5pt}
\end{figure}

Pinterest, a leading content-sharing and social media platform, exemplifies the transformative power of recommendation systems. With billions of Pins and a diverse user base of over 500 million monthly active users \cite{pins}, Pinterest Homefeed, shown in Figure ~\ref{fig:hf}, serves as the primary source of inspiration, providing users with recommendations of high-quality and relevant content. The Homefeed recommendation system uses three stages: retrieval, ranking, and blending. When a user visits the Homefeed, the retrieval stage first retrieves thousands of relevant Pins from billions, based on the user's interests, boards, and other factors. The ranking stage, formulated as a Click-Through Rate (CTR) prediction task, then orders these Pins by predicting their personalized relevance. Finally, the ordered Pins are blended according to business requirements.

While user action history is widely used in CTR models, existing methods often struggle to leverage lifelong user sequences end-to-end for training objectives \cite{transact, SIM}. Serving cost and latency limitations often restrict models to shorter recent user action sequences ($O(10^2)$), neglecting long-term behavioral patterns. Recent attempts to address lifelong user sequences serving either involve compression (losing information) or rely on expensive offline inference and caching \cite{TWIN, TWINv2}. Furthermore, these models typically only encode user actions, lacking the expressive power needed for next-action prediction to improve CTR accuracy.

In this work, we present TransAct V2, a novel model integrating both real-time and lifelong user sequences within the CTR prediction framework.  By employing a new next-action prediction task, TransAct V2 enhances the understanding of user preference and improves recommendation diversity.  Our system design efficiently handles the full lifelong sequences ($O(10^4)$) with low latency, storage, and network cost.  Extensive offline and online experiments validate its effectiveness.
 
TransAct V2 is now serving production traffic on Pinterest’s Homefeed, delivering personalized recommendations to over 500 million users. This deployment not only demonstrates the model’s effectiveness in enhencing user engagement but also showcases its practical and scalable approach to solving real-world application challenges.
Our major contributions are:
\begin{itemize}
\item We introduce TransAct V2, a state-of-the-art model that combines both real-time and lifelong user sequences to improve user engagement and recommendation diversity.
\item We implement a novel next-action loss that empowers the CTR model to better predict user behavior, leading to improved predictive accuracy.
\item We develop web-scale serving solutions for CTR models, ensuring low-latency performance and efficient resource management, accompanied by comprehensive ablation studies highlighting the contribution of each optimization to overall efficiency.
\end{itemize}


\section{Related Work}
\label{sec:related_work}

\subsection{Sequential Recommendation}
Sequential recommendation models primarily aim to capture user interests by analyzing past behavior history. These models are divided into two main categories. The first targets direct prediction of the next item in the user’s sequence, while the second models user sequences within a CTR framework.
Transformer-based architectures have revolutionized sequential recommendation by modeling item dependencies via self-attention ~\cite{SASRec, BERT4Rec, pinnerformer}. While highly effective for pattern identification, they are typically used for pre-training user representations, not end-to-end learning for CTR prediction, and then applied to downstream tasks like item ranking.

In industrial applications, the recommender system usually uses a CTR prediction model in the ranking stage. Recent research has focused on employing sequential transformer encoders to process user action sequences as part of the CTR model~\cite{SIM, UBR4CTR, TWIN, TWINv2, rethink, transact, lirank, hertel2024efficientuserhistorymodeling}
Two-stage user sequence modeling, e.g. SIM \cite{SIM}, UBR4CTR \cite{UBR4CTR}, TransAct \cite{transact}, TWIN \cite{TWIN}, retrieves relevant items then applies self-attention.  However, they either limit sequence length (restricting long-term history exploration) or rely on expensive offline inference and caching.  TWIN v2 ~\cite{TWINv2} attempts to address some challenges by compressing lifelong user action histories using clustering algorithms offline. However, this compression may reduce the sequence's expressive power as critical information can be lost during the compression process.
Furthermore, unlike the category of models that can directly predict the next action, these CTR-focused transformers lack direct next-action prediction capabilities. 

\subsection{Efficient Sequential Model Serving}
Recommendation systems are typically high-throughput systems subject to strict latency constraints. Each request involves ranking thousands of candidates using a sequential model, which presents challenges when handling long user action sequences in real time.  Consequently, many studies employ offline sequence compression techniques, such as clustering methods like TWIN v2 \cite{TWINv2} and Trinity \cite{trinity}, rather than managing long sequences in real time.
Offline sequence compression methods are suboptimal since they are agnostic to the candidate being ranked. Using real-time user sequences allows us to model interaction 
 of the user sequence with the candidate, providing the model with richer information for ranking candidates. 
Notably, TWIN \cite{TWIN} performs compression using a separate model, necessitating real-time model updates. Another recent approach employs generative architectures like HSTU \cite{hstu}, but these typically require models with trillions of parameters to perform well.  Generative recommender (GR) models are also more computationally expensive than regular CTR models.  
Given these considerations, our work concentrates on CTR prediction, as it more effectively addresses the latency and cost constraints typical of industrial-scale applications.

\section{Methodology}
\label{sec:method}
We first describe the ranking model architecture in Section~\ref{sec:ranking model}, then explore lifelong sequence modeling and Next Action Loss. Finally, Section~\ref{sec:system} details TransAct V2's serving and logging system.

\subsection{Preliminary: Ranking Model}
\label{sec:ranking model}
\begin{figure}[h]
  \centering
  \includegraphics[width=0.9\linewidth]{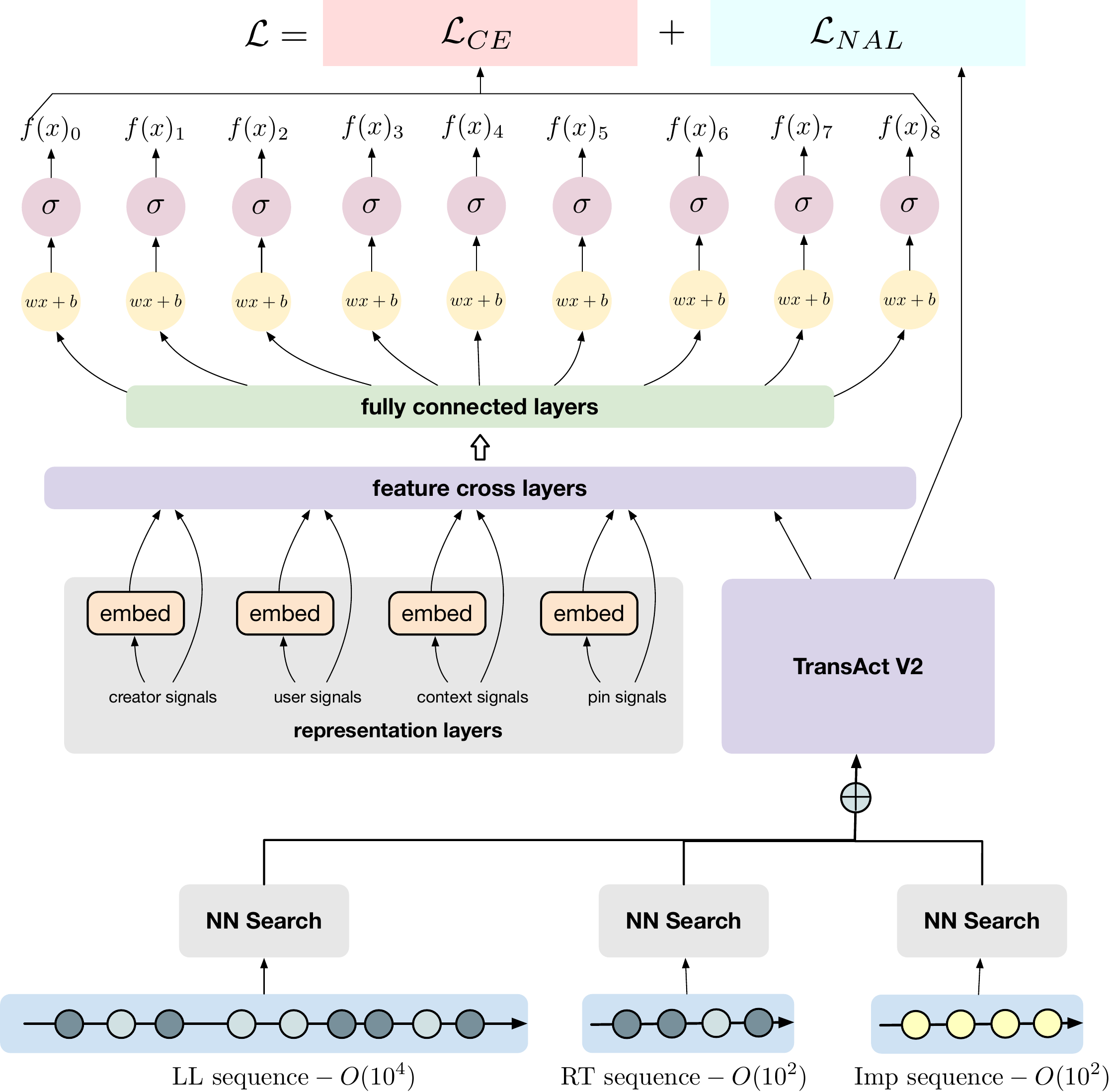}
  \caption{Ranking Model Architecture}
  \label{fig:ranking_model}
  \Description{ranking_model}
\end{figure}

Pinterest's Homefeed ranking model uses a point-wise multi-task learning (MTL) architecture.
As shown in Figure ~\ref{fig:ranking_model}, the model takes context, creator, item, and user features as input to predict the user interaction probability. Similar to existing CTR models~\cite{SIM, transact, rethink, twostage, TWIN}, it employs a standard wide and deep architecture~\cite{cheng2016wide}, encoding various features including user sequences before applying feature interaction layers~\cite{malreddy2024improving} and MLPs to generate head scores.

Each training sample is denoted by the pair \((\vx, \vy)\), where \(\vx\) represents the input feature set, and \(\vy \in \{0, 1\}^{|\vH|}\) is the label vector associated with the action heads in \(\vH\). Each element of \(\vy\) corresponds to the label for a respective action head.
The core loss function of our ranking model is a weighted cross-entropy loss, tailored for optimizing multi-label classification tasks. We define the multi-head prediction loss function as follows:
\begin{equation}
\label{eq:loss}
\mathcal{L}_{CE} = \sum_{h \in H} \left\{ -w_h \left[y_h \log f(\vx)_h + (1 - y_h) \log (1 - f(\vx)_h) \right] \right\}
\end{equation}
Here, \(f(\vx) \in (0, 1)^H\) are the predicted probabilities, with \(f(\vx)_h\) indicating the output probability for head \(h\).  Beyond multi-head prediction, we introduce next-action prediction as an auxiliary task, minimizing the Next Action Loss (NAL), detailed in Section~\ref{sec:NAL}.

\subsection{TransAct V2: Transformer for Lifelong User Action Sequence}
\label{sec:16k}

\begin{figure}[h]
  \centering
  \includegraphics[width=1.0\linewidth]{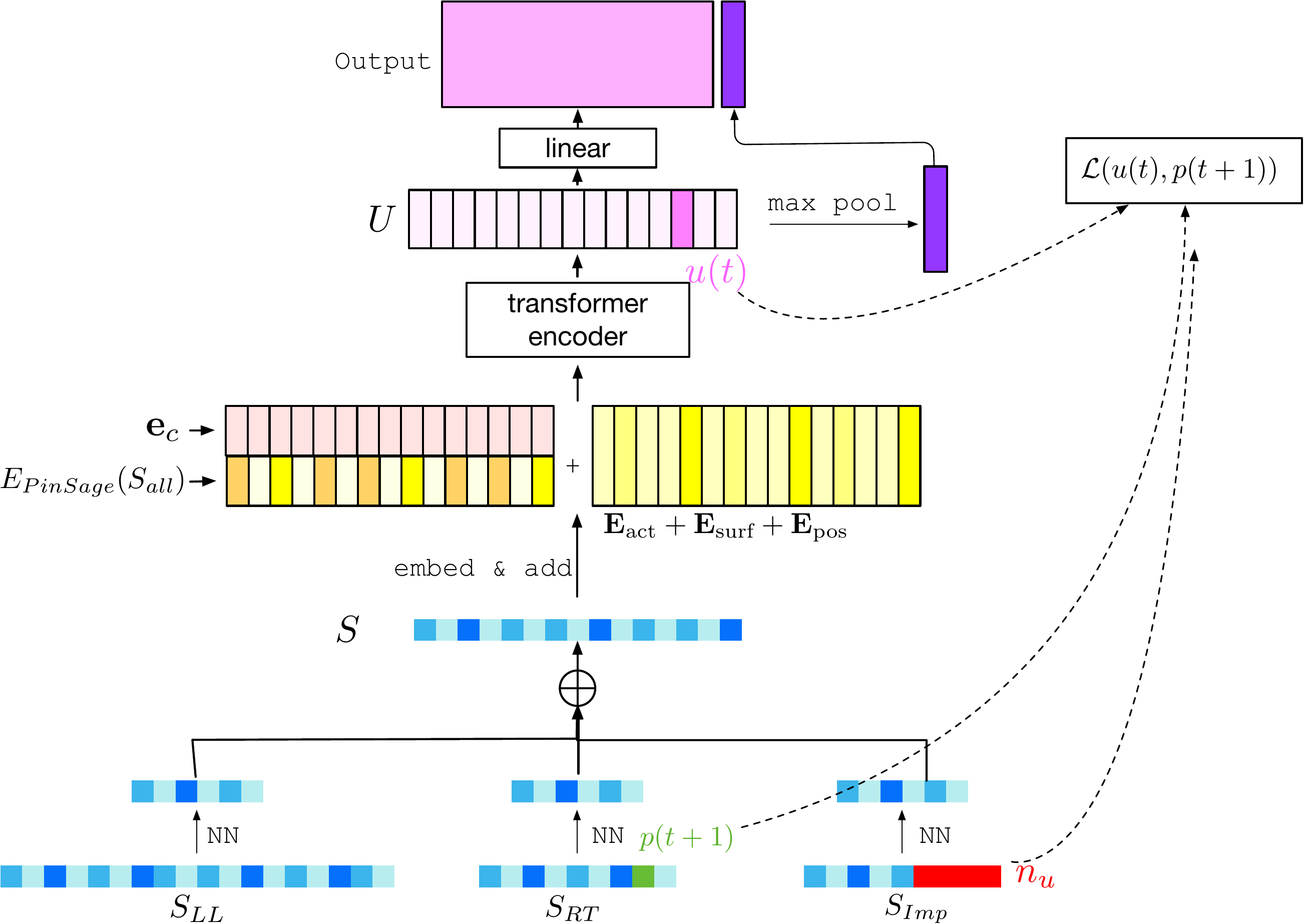}
  \caption{Transact V2 Architecture }
  \label{fig:transact_v2}
  \Description{transact_v2}
\end{figure}
In this work, we introduce TransAct V2, a transformer-based model that integrally combines both real-time and lifelong user action sequences within CTR prediction models to enhance user engagement and recommendation diversity.
Real-time sequences typically capture short-term user behaviors, focusing on their most recent interests. While this is effective for immediate relevance, it often overlooks users' longer-term historical interests, leading to a narrow focus that lacks diversity. This tendency can result in echo-chamber effects~\cite{wang2024itemdissimilaritiesdiversifyingintent, Ge_2020}, where users are repeatedly exposed to similar content, hindering their ability to discover new interests. Such limitations can adversely affect user retention and diminish the overall long-term user experience~\cite{ddp}.
By incorporating lifelong user sequences, our model aims to balance immediate user preferences with historical patterns, fostering a richer and more varied interaction ecosystem. This approach not only improves content diversity but also increases sustained user engagement.

\subsubsection{Lifelong User Sequence Features}
In constructing the lifelong (LL) user sequence $S_{LL}$ for TransAct V2, we focus on capturing meaningful user interactions over an extended period in units of years. The sequence consists of explicit user actions, such as repins\footnote{"Repin" and "save" both refer to saving a Pin to a board on Pinterest.}, clicks, and hides, excluding mere impressions to ensure the relevance of the captured data. 
The maximum length of the LL sequence is chosen based on the 90th percentile of users' past 2 years of action history lengths weighted by visiting frequency. This weighting is crucial as it reflects the behaviors of our most frequent users who significantly contribute to engagement and revenue. 
Each lifelong (LL) sequence token has four features: action timestamp, action type (multi-hot vector if multiple interactions with same pin), action surface (e.g., homefeed, search), and 32-d PinSage embedding~\cite{PinSage} that encapsulates the content of a pin. Action types can be represented as multi-hot vectors when a user interacts multiple times with the same pin. For example, a user might initially perform a close-up view of a pin and subsequently resave it. Real-time sequences ($S_{RT}$) and impression sequences $S_{imp}$ use the same features.
Among these features, the PinSage embedding is the largest data component. To optimize storage and processing, we apply affine quantization to convert the original 32-dimensional \texttt{fp16} PinSage embedding into a 32-dimensional \texttt{int8} vector, halving its size:
$$
q = \text{clamp}\left(\frac{e}{0.65} \times 127, -127, 127\right) \to \texttt{int8}
$$
The scaling of 0.65 ensures precision and minimizes data loss by normalizing the input distribution.
\subsubsection{Modeling Lifelong User Sequence}
\begin{table}[h]
\vspace{-10pt}
\centering
\caption{Table of Notations}
\label{tab:notations}
\renewcommand{\arraystretch}{0.9} 
\begin{tabular}{@{}ll@{}} 
\toprule
\textbf{Notation} & \textbf{Description} \\
\midrule
$\vS_{LL}, \vS_{RT}, \vS_{imp}$ & Lifelong, Real-time, and impression user sequence \\
$\vS_{RT}[:r]$ & Most recent r tokens from $\vS_{RT}$ \\
$\vS_{all}$ & Concatenated sequence from all sources \\
$\mathbf{E}_{\text{act}}, \mathbf{E}_{\text{surf}}$ & Embedding matrix for user action, surface \\
$\mathbf{E}_{PinSage}(\vS)$ & PinSage embedding matrix of sequence $\vS$ \\
$\mathbf{E}_{\text{pos}}$ & Positional encoding matrix \\
$e_c$ & PinSage embedding of candidate pin \\
$u(t)$ & User embedding at time $t$ \\
$p_u(t)$ & Positive sample PinSage embedding at $t$ \\
$n_u$ & Negative samples for user $u$ \\
\bottomrule
\end{tabular}
\end{table}

\textbf{Nearest Neighbor Search}. As shown in Figure~\ref{fig:transact_v2}, we process both the real-time and lifelong user sequences using the same transformer encoder. Following an approach similar to TransAct~\cite{transact}, we begin by using the candidate item, denoted as $c$, as an anchor to perform nearest neighbor (NN) searches on three distinct sequences: the lifelong sequence $\vS_{LL}$, the real-time sequence $\vS_{RT}$, and the impression sequence $\vS_{imp}$. In addition, we always keep the most recent $r$ actions $\vS_{RT}[:r]$ to ensure that the model consumes user's most fresh actions regardless of the similarity with the candidate item.

The NN search results in sub-sequences from each $\vS_{LL}$, $\vS_{RT}$, and $\vS_{imp}$, which are concatenated along with $\vS_{RT}[:r]$ as follows:
\begin{equation}
    \vS_{all} = \text{NN}(\vS_{LL}, c) \oplus \vS_{RT}[:r] \oplus \text{NN}(\vS_{RT[r:]}, c) \oplus \text{NN}(\vS_{imp}, c)
    \label{equation: nn_combination}
\end{equation}
where $\text{NN}(\vS, c)$ is computed by calculating the dot product between the candidate PinSage embedding $\mathbf{e}_c$ and the PinSage embeddings of sequence $\vS$, $\mathbf{E}_{\text{PinSage}}(\vS)$, then selecting the top $K$ elements based on similarity:   
\begin{equation}  
    \text{NN}(\vS, c) = \{\mathbf{\vS}_i | i \in \text{argsort}(\mathbf{E}_{\text{PinSage}}(\vS) \cdot \mathbf{e}_c)[-K:]\}  
    \label{equation: nn_detail}  
\end{equation}  
It is important to note that $\vS_{LL}$ has a length of $O(10^4)$, whereas both $\vS_{RT}$ and $\vS_{imp}$ are much shorter, with lengths of $O(10^2)$.
This final sequence $\vS_{all}$ is constrained to several hundred elements ($O(10^2)$), ensuring that the computational cost of processing through the transformer encoder remains manageable. 

\textbf{Feature Encoding}:
To incorporate the metadata of each token in user sequences (action type, surface), we use trainable embedding tables to project them to low-dimensional vectors. The action type sequence is projected to an embedding matrix $\mathbf{E}_{\text{act}} \in \mathbb{R} ^{|S| \times d_{act}}$, where $d_{act}$ is the dimension of the action type embedding.
Similarly, the surface feature is embedded as $\mathbf{E}_{\text{surf}} \in \mathbb{R} ^{|S| \times d_{surf}}$.
Positional encoding is implemented as a learnable parameter $\mathbf{E}_{\text{pos}} \in \mathbb{R} ^{|S| \times d_{pos}}$, allowing the model to adaptively learn positional information within the sequence. 
For each token in $S_{all}$, we concatenate the candidate pin's PinSage embedding $\mathbf{e}_c$ as our early fusion approach~\cite{transact}.
To make all components additive, we set $d = d_{\text{act}} = d_{\text{surf}} = d_{\text{pos}} = 2d_{\text{PinSage}}$. The final encoded user action sequence is formed by adding all the above components.
\begin{equation}
    \vF = \texttt{CONCAT}(\mathbf{E}_{\text{PinSage}}(\vS_{all}), \mathbf{e}_c) + \mathbf{E}_{\text{act}} + \mathbf{E}_{\text{surf}} + \mathbf{E}_{\text{pos}} \in \mathbb{R}^{|S| \times d}
    \label{equation: feature_encoding}
\end{equation}

\textbf{Transformer Encoder}:
$\vF$ is processed using a transformer encoder composed of two layers, with each layer utilizing a single attention head. The model dimensionality is set to 64, and the feed-forward network within each layer has a dimension of 32. Note that a causal mask is applied to the transformer encoder. We provide more details about the hyper-parameter choice in \ref{sec:tfmr_hpt}.
The transformer encoder output $\mU = (u(0):u(|S|-1))\in \mathbb{R} ^ {|S| \times 2d}$ is used by two downstream tasks.
The first one is a multi-head prediction task. As shown in Figure~\ref{fig:transact_v2}, a linear layer and a max pooling layer are applied on top of $\mU$ to produce the output. The output will be fed into the feature crossing layers in Figure~\ref{fig:ranking_model} for multi-head prediction. The second task is the next action prediction task as discussed in next section.

\subsection{Next Action Loss}
\label{sec:NAL}
We have discussed how the model summarizes user action history. To fully leverage this user sequence information, we designed a separate next-action prediction task, training the transformer to anticipate a user's next action. Our experimental results demonstrate that this auxiliary task alongside multi-head prediction significantly enhances ranking performance.

\subsubsection{Next Action Prediction Task}
Specifically, for a given user's most recent actions from $S_{RT}$, we predict whether the engagement of the $(t+1)$-th pin is positive or negative using the transformer-processed user embedding at time $t$. The prediction loss is constructed in a contrastive loss manner. For a given user $u$, denote the $t$-th column in the output of the transformer $\mU$ by $u(t)$, which can also be viewed as the embedding of the user $u$ at time $t$ when causal masking is applied. We use $p_u(t+1)$ to denote the PinSage ~\cite{PinSage} embeddings for the positive sample at $t+1$, and $n_u$  to denote the negative samples for user $u$. The specifics of positive/negative sample selection will be discussed in the next section. 
The next action loss (NAL), denoted as $\mathcal{L}_{NAL}$, employs a sampled softmax (SSM) loss function. SSM is a common technique in contrastive learning, and has been applied in recommendation systems for learning user representations at Pinterest ~\citep{pinnerformer}. 
\begin{equation}
    \mathcal{L}(u(t),p(t+1)) = -\log\left(\frac{e^{\langle u(t), p_u(t+1) \rangle}}{e^{\langle u(t), p_u(t+1) \rangle} + \sum\limits_{n_u} e^{\langle u(t), n_u \rangle}}\right),
    \label{equation: SSM}
\end{equation}
\begin{equation}
    \mathcal{L}_{NAL} = \sum_u\sum_t\mathcal{L}(u(t),p_u(t+1)),
    \label{equation: NAL}
\end{equation}
where $N$ is the number of negative samples and $\langle\cdot,\cdot\rangle$ is for vector inner product. 
$\mathcal{L}_{NAL}$ is then weighted, and summed with the multi-head cross entropy loss  $\mathcal{L}_{CE}$ during training. 
\begin{equation}
    \mathcal{L} = \mathcal{L}_{CE} + w_{NAL} * \mathcal{L}_{NAL}
    \label{equation: loss}
\end{equation}
\subsubsection{Key Modeling Design of NAL}
Unlike standard self-attention, which allows all tokens to attend to each other, a \textbf{causal\ mask} is used to restrict tokens to attend only to preceding ones, thereby preventing information leakage during the next action prediction task. For instance, when predicting the $t$-th action, the model has access only to actions from 0 to $t-1$.

The choice of loss function for the next action prediction is another important aspect. While cross-entropy is a popular choice for classification, we utilize \textbf{sampled softmax loss} because it offers greater flexibility in adjusting the ratio of positive to negative samples in Equation~\ref{equation: SSM}, leading to consistently better performance.

In terms of sample selection, for positive samples, we utilize all tokens from $S_{RT}[:r]$ that exhibit positive engagement types. For negative samples, we explored two methods. The first, in-batch random negative sampling, involves selecting a user $u_j$ randomly from the same batch as user $u_i$ and choosing $N$ Pins from $u_j$'s sequence. This assumes that different users within the same batch typically have different engagement interests. The second, impression-based negative sampling, selects Pins from $S_{imp}$ that the current user has viewed but not engaged with further, suggesting low interest.
Our findings, detailed in Section~\ref{sec:nal_ablation}, reveal that \textbf{impression-based negative samples} are more effective, providing a more challenging set that results in improved ranking performance. 


\subsection{Serving and Logging System Design}
\label{sec:system}

One of the major challenges of utilizing a LL sequence in CTR models lies in the significant \textbf{serving} and \textbf{storage costs} that come with the extended sequence length. 
Let $L$ represent the length of $S_{LL}$ and $N$ denote the average number of items per ranking request. Under these conditions:
\begin{itemize}
    \item The corresponding \textbf{feature storage cost} scales as $\mathcal{O}(L)$.
    \item The \textbf{network cost} scales as $\mathcal{O}(NL)$.
\end{itemize}

\begin{figure}[h]
  \centering
  \includegraphics[width=0.8\linewidth]{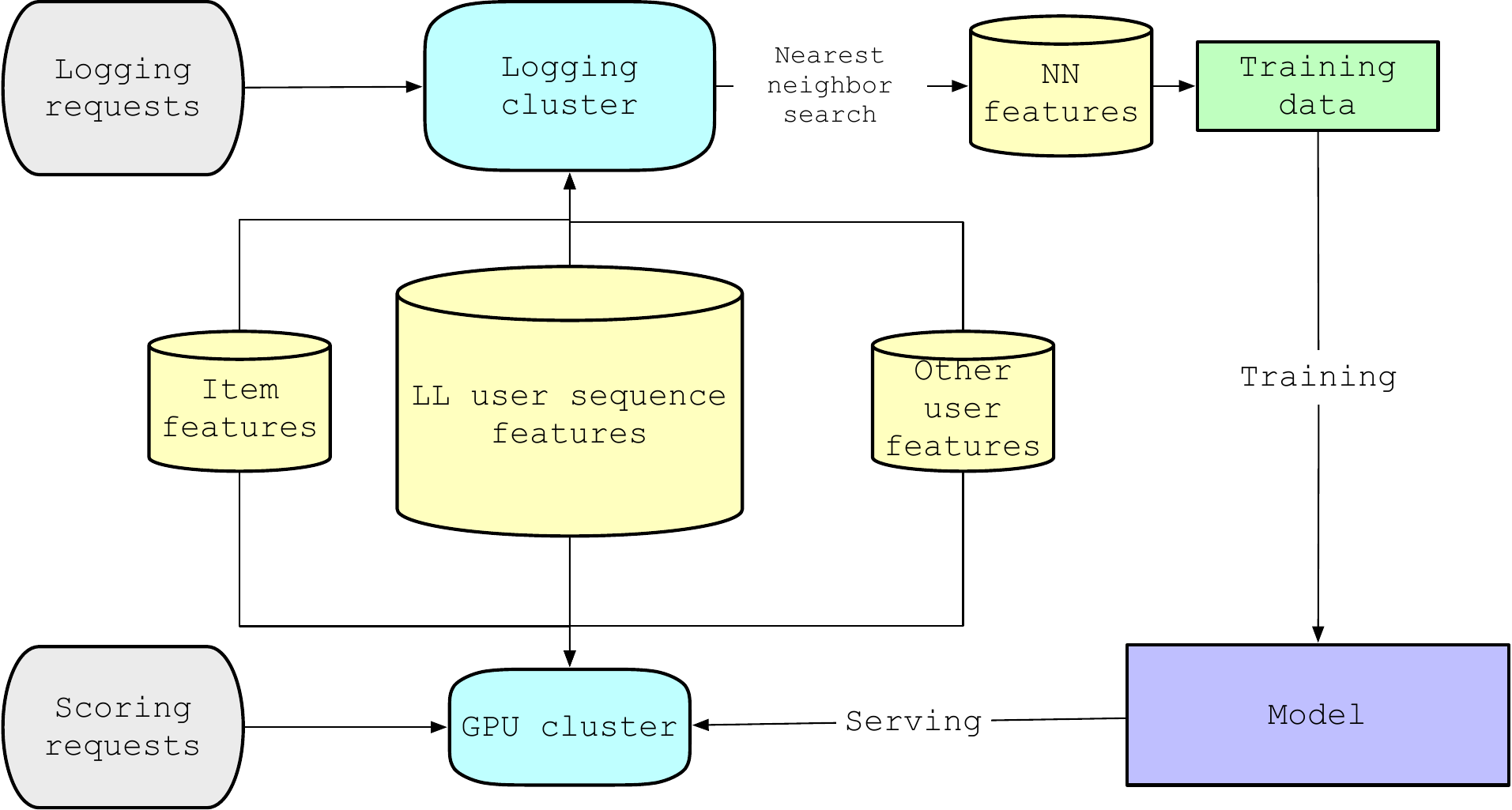}
  \caption{An illustration of the Online Serving System}
  \label{fig:system_design}
  \Description{system_design}
  \vspace{-10pt}
\end{figure}

In this work, $L$ is at the scale of $\mathcal{O}(10^4)$, introducing substantial challenges in both model serving and network efficiency. Without any optimizations, serving the LL sequence would make the system prohibitively expensive and inefficient.
To address both serving latency and storage bottlenecks, we propose a new machine learning data pipeline specifically for super long user sequence models, as illustrated in Figure~\ref{fig:system_design}.

These optimizations ensure that our system can handle significantly longer user sequences without compromising latency, storage efficiency, or model performance.

\subsubsection{Data Pipelines}
All features utilized by the ranking model are usually stored in cache at serving time and subsequently logged into the training data. However, LL sequence features are too large for efficient caching and network transfer. To address the resulting network and storage costs, we implement a \textbf{nearest neighbor (NN) feature logging strategy} within the data pipeline, as depicted in Figure~\ref{fig:system_design}. Rather than transferring and storing the entire sequence of length $L$, we perform the NN search detailed in Equation~\ref{equation: nn_detail} and log only the relevant NN features to the training dataset. This approach reduces data storage complexity from 
$\mathcal{O}(L)$ to $\mathcal{O}(1)$, thereby optimizing resource utilization significantly.



Our framework employs distinct data processing pipelines for \textbf{training and serving}, optimized for efficiency in both scenarios. 
During training, the model consumes pre-computed nearest neighbor (NN) features from logged data, significantly reducing data loading overhead. These NN features are passed directly into the transformer encoder.
In contrast, during serving, when receiving a ranking request, the system retrieves the full LL sequence of the target user. When the user features and the list of $N$ items are sent to model inference, we broadcast the LL sequence features across all items, followed by an on-device nearest neighbor search to extract relevant NN features. These features are then fed into the transformer encoder for inference.
This design balances storage efficiency, computational cost, and alignment between the training and serving stages, ensuring scalability in production environments.

\begin{itemize}
    \item \textbf{Storage Efficiency:} Logging NN features instead of full sequences minimize storage and bandwidth costs.
    \item \textbf{Training-Serving Alignment:} Although the feature processing paths differ, the NN feature representation remains consistent.
    \item \textbf{Inference Scalability:} Real-time NN search on-device ensures scalability and low-latency predictions in industrial deployment environments.
\end{itemize}

\subsubsection{Serving Optimizations}  \label{optimizations}
To serve LL user sequences in real-time high-QPS systems without incurring huge costs, we jointly optimize the inference server and the model to reduce latency.

Pinterest's recommender models use a custom C++ inference engine and CUDAGraph~\cite{cudagraph} to reduce the CUDA kernel launch overhead as detailed in TransAct~\cite{transact}. Requests are dynamically batched into mini-batches, prepared on pageable CPU memory, copied into a pre-allocated pinned memory buffer (per CUDAGraph instance), and finally copied from the pinned memory buffer to static GPU memory locations for the model forward pass.

OpenAI’s Triton framework~\cite{triton,triton2} has recently emerged as a powerful tool for writing custom kernels to improve model efficiency. It has been used widely in LLM training as well as recommendation system~\cite{cce,liger,jaggedattn}. Triton allows more fine-grained control over tensor tiling and GPU L2 Cache (Shared Memory or SRAM), enabling smarter caching and in-place fused operations. In our work, we adopt Triton to build custom transformer kernels to optimize model serving.
In addition, processing $\mathcal{O}(NL)$-sized tensors can lead to substantial CPU-to-GPU and intra-GPU data transfers. By fusing expensive operations into custom kernels, we reduce both the number of kernel calls and the need for intermediate data transfers, thus improving overall serving efficiency.

\begin{figure}[h]
\vspace{-8pt}
  \centering
  \includegraphics[width=0.97\linewidth]{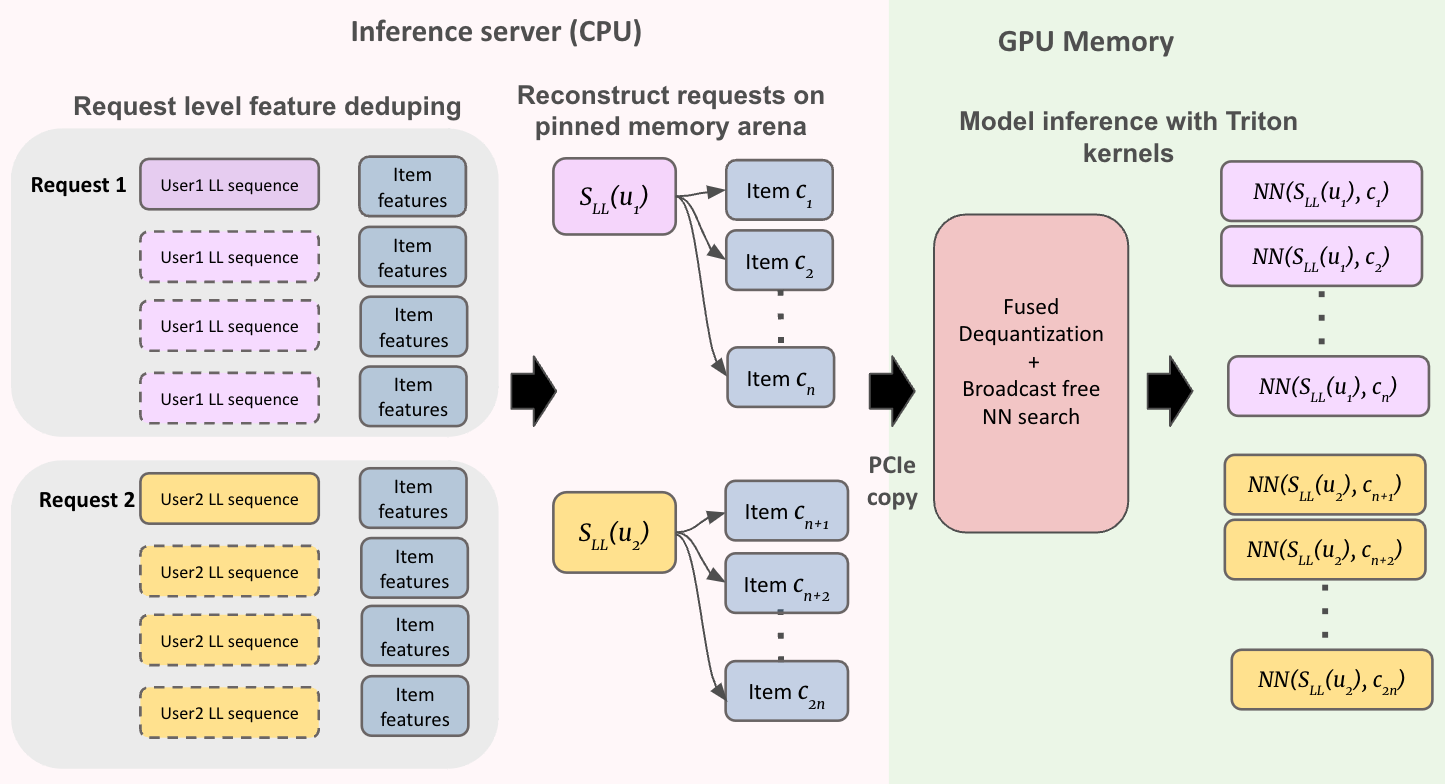}
  \caption{Serving setup with server side optimizations and custom sparse tensor implementation for request level de-duplication}
  \label{fig:bcast}
  \Description{bcast}
  \vspace{-8pt}
\end{figure}

\textbf{Request Level De-duplication.} 
In a typical ranking setup, a single inference request ranks 
$N$ items for one user. Request-level features like user sequences are replicated (broadcasted) to each item on the CPU and then transferred to the GPU for inference. The LL sequence features constitute a significant portion of the total feature size. Instead of broadcasting these request-level sequence features, we introduce a new sparse tensor format that stores de-duplicated request-level features and offsets. As shown in Figure \ref{fig:bcast}, this format supplies all necessary metadata for the model to handle request-level sequences without broadcasting. Furthermore, we developed a custom Triton kernel to conduct an NN search (Eq~\ref{equation: nn_detail}) directly on the de-duplicated requests (broadcast free NN search), eliminating the need to broadcast request-level sequence features on the GPU. This optimization yields an 8x reduction in PCIe data transfers for sequence features. 

\textbf{Fused Sequence Dequantization. } As discussed in Section~\ref{sec:16k}, the PinSage sequence feature is stored in a quantized int8 format. Before NN search, this tensor must be dequantized to \texttt{float16} and normalized. In PyTorch, L2 normalization typically involves launching four separate kernels of squaring, summation, clamping, and division. Each kernel processes this sizable tensor, and at high QPS, these operations can become a performance bottleneck. Our fused implementation combines this with the NN search kernel resulting in a 20\% reduction in model latency compared to the native PyTorch implementation.

\textbf{Single Kernel Unified Transformer (SKUT). }  Leveraging our model's low dimensionality ($d_{model} = 64$), we developed a novel fused transformer that achieves a forward pass performance that is 6.6 times faster than PyTorch. 
The transformer is implemented as a single merged custom Triton kernel. The strong latency improvements are achieved mainly by only materializing the QKV sequence tensors on-the-fly during tiled attention. The feed-forward and layer-norm operations are fused immediately after to produce the final output tile.

Our inference server utilizes Nvidia A10G GPUs (8 MB shared memory) to process Transformer inputs.  Given the small dimensionality, we can accommodate all transformer weights — the three projection tensors $Q$, $K$, $V$ of shape (64, 64) and the feed-forward network weights $W_1$, $W_2^T$ of shape (64, 32) alongside the input tensor within 6 MB of shared memory.  This allows us to fuse all transformer operations - QKV projection, Flash Attention, layer normalization, and the feed-forward network, into a single fused operation directly on SRAM, significantly reducing GPU transfer overhead and kernel calls.

\textbf{Pinned Memory Arena.}
Memory Arena is a thread-local, contiguous block of pre-allocated memory, which is reset and reused across subsequent inference requests. Pre-allocation of memory ensures that no tensor memory is allocated for inference on the fly.  
During inference, each mini-batch of size 128 results in a 64 MB payload. This large payload introduces two copy bottlenecks - Pageable to Pinned memory copy and Pinned CPU memory to GPU copy. The latter is being sufficiently addressed by request-level deduplication. To compensate for the increased memory pinning cost incurred from this new feature, we implemented a Pinned Memory Arena.
Our implementation backed the Memory Arena with pinned memory and constructed collated mini-batches directly onto it. This completely eliminated the pageable to pinned copy step and improved inference speed for the final experiment setup by up to 35\%.

\section{Experiment}
\label{sec:exp}
In this section, we will present extensive offline and online A/B experimental results of TransAct V2. We compare the performance of TransAct V2 with baseline models using Pinterest's internal training data.
\subsection{Experiment Setup}
\subsubsection{Dataset}
Our offline training dataset is gathered and downsampled from two weeks of user activities on Pinterest Homefeed. The training data contains 6.9 billion training instances of 182 million users and 350 million Pins.
The model is trained from scratch using randomly initialized weights and evaluated on an evaluation dataset collected 7 days after the last day of training data. 

In this paper, we conduct all experiments with the Pinterest dataset. Our experiments require lifelong action sequence, real-time action sequence and impression sequence along with metadata features, such as item embeddings, action types, surface types.  
To the best of our knowledge, publicly available CTR prediction datasets lack these features, though they are readily reproducible in industry recommender systems.


\subsection{Offline Experiment}

\subsubsection{Metrics}

Our model performance is evaluated using the HIT@3 metric. A "chunk" of recommendations, denoted as $\vc = [p_1, p_2, \dots, p_n]$, represents a set of Pins recommended to a user simultaneously. Each evaluation data instance for the ranking model includes a user identifier $u$, a pin identifier $p$, and a chunk identifier $c$.
For evaluation, outputs are grouped by user and chunk identifiers $(u, c)$ to aggregate results from the same ranking request. Within each group, Pins are sorted based on a final ranking score $\mathcal{S}$, a linear combination of the outputs from the ranking model's heads, denoted by $f(\vx)$:
$$
\mathcal{S} = \sum_{h \in \mathcal{H}} w_h f(\vx)_h
$$
Here, $\mathcal{H}$ represents the set of model heads, and $w_h$ is the weight applied to each head's output. The top $K$ Pins from each chunk are considered, and HIT@K is computed, noted as $\gamma_{c, h}$, which counts the number of top-K Pins bearing a label of 1 for head $h$. For example, if chunk $\vc = [p_1, p_2, p_3, \dots, p_n]$ is sorted by $\mathcal{S}$, and the user repins $p_1$ and $p_4$, then $\text{HIT}@K$ for repin is $\gamma_{c, \text{repin}} = 1$ when $K = 3$.
The aggregated HIT@3 for each head $h$ is calculated  over all users $U$ and their corresponding chunks $C_u$ as follows:
$$
\text{HIT@3}/h = \frac{\sum_{u \in U}\sum_{c \in C_u} \gamma_{c, h}}{|U|}
$$
For positive engagement (e.g., repins, clicks), higher HIT@K is better.  For negative engagement like hides, lower $\text{HIT@K}/\text{hide}$ is preferable.


\subsubsection{Results}

\begin{table}
\centering
\caption{Offline evaluation comparing existing methods with ours. ($^*$ statistically insignificant)} \label{tab:offline_res}
\resizebox{.45\textwidth}{!}{%
\begin{tabular}{lcc}
\toprule
{Methods} &  HIT@3/repin & HIT@3/hide\\
\midrule
BST (RT sequence)~\cite{alibaba_seq_tfmr} & +6.04\% & -0.49\%* \\
TransAct (RT sequence)~\cite{transact} & +7.74\% & -6.86\% \\
TransAct (RT sequence) + NAL\textsubscript{in\_batch} & +8.41\% & -8.63\% \\
TransAct (RT sequence) + NAL\textsubscript{imp} & +8.92\% & -9.08\% \\
TransAct V2 (RT seq + LL seq + NAL\textsubscript{imp}) & \textbf{+13.31\%} & \textbf{-11.25\%} \\
\bottomrule
\end{tabular}%
}
\end{table}

We compare our proposed method with existing models for CTR prediction that utilize user sequence features. We first compare with the Wide and Deep Learning model~\cite{cheng2016wide} without incorporating user sequence features. Additionally, we compare our model with BST~\cite{alibaba_seq_tfmr} and TransAct~\cite{transact}, which both focus on real-time sequence modeling. TWIN\cite{TWIN} is not compared in this work because it relies on a computationally expensive offline inference system.
Models such as BERT4Rec~\cite{BERT4Rec} are omitted from direct comparison due to differences in problem formulation.

As presented in Table~\ref{tab:offline_res}, TransAct V2, which incorporates both RT and LL sequences along with impression-based NAL, achieves superior performance, outperforming all other approaches.  It improves HIT@3/repin by 13.31\% and reduces HIT@3/hide by 11.25\%, demonstrating its ability to enhance positive user engagement and minimize negative interactions.
Impression-based Negative Sampling (NAL\textsubscript{imp}) is crucial to this improvement. 
By incorporating these impression-based samples, the model benefits from a more nuanced understanding of user disinterest, leading to better calibration of the ranking model. As a result, our experiments confirm that NAL\textsubscript{imp} along with the lifelong user sequence delivers significant improvements in ranking effectiveness.
\subsection{Online Experiment}
In the online experiment, we deployed the models in offline experiments.  Each group served 1.5\% of the Homefeed page visitors.  To protect user experience, we used a strong baseline — \textbf{TransAct (RT sequence)} from Table~\ref{tab:offline_res}. 

\subsubsection{Metrics}

Key Homefeed metrics include \textbf{Homefeed Repin Volume} (higher is better, correlating with HIT@3/repin offline) and \textbf{Homefeed Hide Volume} (lower is better, indicating irrelevant recommendations and poor user experience).  We also consider \textbf{Impression Diversity} (higher is better, reflecting broader content exposure).

\subsubsection{Online Results}

Table~\ref{tab:online_metric} shows the online performance of our model. We observe that the $NAL_{imp}$ performs particularly well in reducing the Homefeed Hide Volume by 6.26\%. This indicates that using impression-based negative sampling effectively filters out irrelevant recommendations, leading to a better user experience by presenting more pertinent content.
TransAct V2 integrated with $NAL_{imp}$ achieves the best overall performance. Specifically, it led to a significant increase of 6.35\% in Homefeed Repin Volume, suggesting that the model provides highly relevant recommendations to users. Additionally, it reduces Homefeed Hide Volume by 12.80\%, further demonstrating the model's capability in curating content that users find desirable. 
Moreover, the model also improves Impression Diversity, enhancing the variety of content presented to users and enriching their browsing experience. 


\begin{table}
  \caption{Online Evaluation Metrics compared with TransAct as baseline. ($^*$ statistically insignificant)}
  \label{tab:online_metric}
  \centering
  
  \begin{tabular}{lrr}
    \toprule
    Online Metrics & $NAL_{imp}$ & TransAct V2 \\
    \midrule
    Homefeed Repin Volume & +0.27\%* & +6.35\% \\
    Homefeed Hide Volume & -6.26\% & -12.80\% \\
    Impression Diversity & +0.08\%* & +0.45\% \\
    Time Spent on App & +0.10\%* & +1.41\% \\
    \bottomrule
  \end{tabular}
    \vspace{0.5em} 
  \begin{flushleft}
  \small Note: 1\% increase in repin volume is considered as a substantial gain.
  \end{flushleft}
\end{table}

\subsection{Ablation Study}
We first evaluate design choices for Next Action Loss and the transformer encoder.  Then, we examine different components of serving optimizations.

\subsubsection{Next Action Loss Ablation}
\label{sec:nal_ablation}

We present the ablation study results for NAL design choices. 
Impression-based negative sampling significantly outperforms in-batch sampling  as shown in Table~\ref{table:nal_neg_samples}.  This is because impressions provide stronger and more challenging negatives, improving model generalization and, crucially, enhancing model accuracy in predicting negative outcomes and reducing hides.  By training the transformer to recognize and respond to more nuanced negative signals, the model achieves a more comprehensive understanding of user preferences.
Additional NAL ablations are provided in appendix.

\begin{table}
\caption{Offline evaluation of NAL negative sample selection.}
\centering
\resizebox{0.4\textwidth}{!}{
    \begin{tabular}{ccccc}
        \toprule
        Negative sample selection & HIT@3/repin & HIT@3/hide \\
        \midrule
        In-batch neg samples & +0.63\% & -1.90\%\\
        Impression-based neg samples & \textbf{+1.10\%} & \textbf{-2.39\%}\\
        \bottomrule
    \end{tabular}%
}
\label{table:nal_neg_samples}
\end{table}

\subsubsection{Transformer Hyperparameters}
\label{sec:tfmr_hpt}
\begin{figure}[h]
  \vspace{-8pt}
  \centering
  \includegraphics[width=0.8\linewidth]{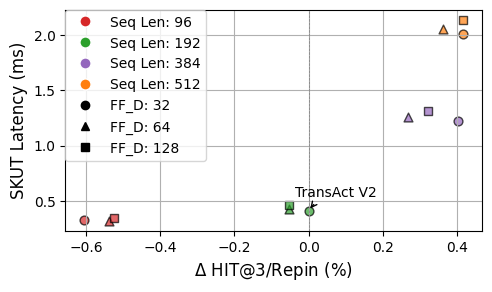}
  \caption{Model Performance and inference latency trade-offs on different transformer hyper-parameter settings}
  \label{fig:tfmr_hpt}
  \Description{tfmr_hpt}
  \vspace{-10pt}
\end{figure}

We conducted extensive tuning of transformer hyperparameters, specifically focusing on the feed-forward dimension and input sequence length $|S_{all}|$. 
Our primary goal is to determine the configuration that would deliver optimal performance, balancing both repin metrics and inference costs/latency.
Comprehensive evaluation identified the optimal configuration: sequence length 192, 2 layers and feed-forward dimension 32. This balanced strong repin metrics with low inference latency.  Figure~\ref{fig:tfmr_hpt} shows that increasing sequence length and feed-forward dimension improves HIT@3/repin but significantly increases inference latency.

\subsubsection{Serving Optimization Ablation}

\label{sec:serving_ablation}

\textbf{Single Kernel Unified Transformer} -   
SKUT performance was evaluated by measuring forward latency across sequence lengths 64-2048.  
Using PyTorch's memory-efficient implementation~\cite{memeffattn} as the baseline, 
SKUT achieved 85.09\% lower latency and 13.24\% less GPU memory utilization in our production setting (Batch size 256, Sequence length 192).
Table~\ref{tab:latency-memory-decrease} shows SKUT's advantage over baseline implementation across different sequence length. SKUT also outperforms baseline across different batch sizes (see Appendix~\ref{appendix_skut} for details).
These gains primarily come from avoiding Q, K, V tensor materialization. The peak latency reduction occurs at a sequence length of 192 due to kernel tuning for production input shapes. SKUT also maintains its latency advantage over PyTorch's implementation with for longer sequences.
It is noteworthy that FlashAttention-2~\cite{fav2} was not directly compared because it lacks support for the custom masks required for masking tokens for NAL and padding. However, we still provide comparison without masks in Appendix\ref{appendix_skut}.


\begin{table}[h]
\vspace{-8pt}
    \centering
    \caption{Latency and Memory Reduction (SKUT vs. PyTorch)}
    \footnotesize
    \setlength{\tabcolsep}{4pt} 
    \renewcommand{\arraystretch}{0.9} 
    \begin{tabular}{|c|c|c|}
        \hline
        \textbf{Seq Length} & \textbf{Latency Decrease} & \textbf{Memory Decrease} \\
        \hline
        64   & -59.29\% & -19.60\% \\
        128  & -81.58\% & -10.06\% \\
        \textbf{192} & \textbf{-85.09\%} & \textbf{-13.24\%} \\
        256  & -84.79\% & -15.42\% \\
        512  & -82.27\% & -18.12\% \\
        1024 & -74.85\% & -15.22\% \\
        2048 & -73.10\% & -9.80\% \\
        \hline
    \end{tabular}
    \vspace{-8pt}
    \label{tab:latency-memory-decrease}
\end{table}

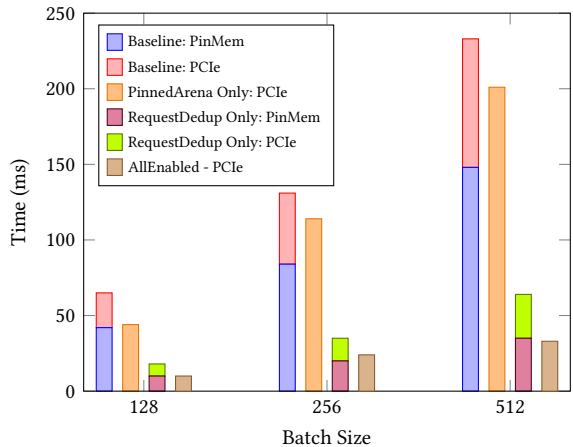
\begin{figure}    
\vspace{-2pt}
    \centering    
    \begin{tikzpicture}[               
            every axis/.style={
                width=0.95\linewidth,    
                height=0.78\linewidth,    
                ybar stacked,    
                ymin=0,ymax=250,  
                symbolic x coords={128, 256, 512},    
                xtick=data,    
                xlabel={\small{Batch Size}},    
                ylabel={\small{Time (ms)}},    
                ticklabel style={font=\small},
                enlarge x limits={abs=0.8cm},    
                bar width=6pt,   
                legend image code/.code={
                \draw [#1] (0cm,-0.1cm) rectangle (0.2cm,0.12cm); },
                legend style={at={(1,-0.2)}, legend columns=1,   
                          /tikz/every even column/.append style={column sep=0.14cm},  
                          font=\scriptsize},
                          legend pos = north west,
                          legend cell align=left
            }  
        ]  
        \begin{axis}[bar shift=-15pt,hide axis]    
        \addplot coordinates {(128,42) (256,84) (512,148)};   
        \label{BaselinePinMem}
        \addplot coordinates {(128,23) (256,47) (512,85)}; 
        \label{BaselinePCIe}
        
    \end{axis}
    \begin{axis}[bar shift=-5pt,hide axis]    
        \addplot [orange!90!black, fill=orange!50!white] coordinates {(128,44) (256,114) (512,201)};      
        \label{PinnedArenaPinMem}
    \end{axis}    
    
    \begin{axis}[bar shift=5pt,hide axis]    
        \addplot [purple!40!black, fill=purple!50!white] coordinates {(128,10) (256,20) (512,35)};       \label{RequestDedupPinMem}
        \addplot [lime!45!black, fill=lime]     coordinates {(128,8) (256,15) (512,29)};      
        \label{RequestDedupPCIe}
    \end{axis}    
    
    \begin{axis}[bar shift=15pt]    
        \addlegendimage{/pgfplots/refstyle=BaselinePinMem}\addlegendentry{Baseline: PinMem}
        \addlegendimage{/pgfplots/refstyle=BaselinePCIe}\addlegendentry{Baseline: PCIe}
        \addlegendimage{/pgfplots/refstyle=PinnedArenaPinMem}\addlegendentry{PinnedArena Only: PCIe}
               
        \addlegendimage{/pgfplots/refstyle=RequestDedupPinMem}\addlegendentry{RequestDedup Only: PinMem}
        \addlegendimage{/pgfplots/refstyle=RequestDedupPCIe}\addlegendentry{RequestDedup Only: PCIe}

        \addplot [brown!70!black, fill=brown!60!white]  coordinates {(128,10) (256,24) (512,33)}; 
        \addlegendentry{AllEnabled - PCIe}
    \end{axis}    
  

    \end{tikzpicture}    
  
    \caption{Copy Time components by batch size and configurations at P99 latency.}    
    \vspace{-8pt}
    \label{fig:copy-time}  
\end{figure}

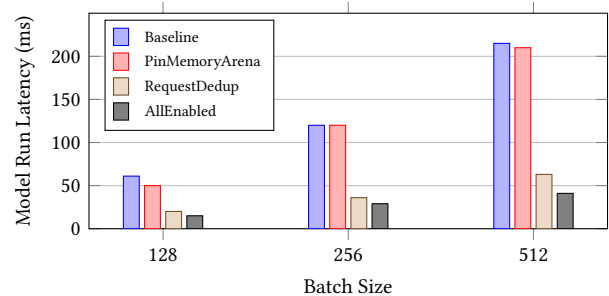
\begin{figure}[h]
\vspace{-2pt}
    \centering
    \begin{tikzpicture}
        \begin{axis}[
            ybar,
            width=1.0\linewidth,
            height=0.25\textwidth,
            bar width=6pt,
            ylabel={\small{Model Run Latency (ms)}},
            xlabel={\small{Batch Size}},
            symbolic x coords={128, 256, 512},
            xtick=data, 
            ytick={0,50,100,150,200},
            ticklabel style={font=\small},
            legend image code/.code={
            \draw [#1] (0cm,-0.1cm) rectangle (0.2cm,0.13cm); },
            legend style={legend columns=1, font=\scriptsize,column sep=0.14cm},
            legend pos=north west,
            legend cell align=left,
            ymajorgrids=true,
            ymin=0,
            ymax=250,
            enlarge x limits=0.2,
            title style={font=\small}
        ]
            \addplot coordinates {(128, 61) (256, 120) (512, 215)};
            \addlegendentry{Baseline}
            \addplot coordinates {(128, 50) (256, 120) (512, 210)};
            \addlegendentry{PinMemoryArena}
            \addplot coordinates {(128, 20) (256, 36) (512, 63)};
            \addlegendentry{RequestDedup}
            \addplot coordinates {(128, 15) (256, 29) (512, 41)};
            \addlegendentry{AllEnabled}
        \end{axis}
    \end{tikzpicture}
    \caption{P99 Latency comparison across techniques for various batch sizes.}
    \vspace{-8pt}
    \label{fig:model-runlatency}
\end{figure}

\textbf{Server Optimizations}. Results in Figures \ref{fig:copy-time}, \ref{fig:model-runlatency}, and \ref{fig:heatmap-final-reducton} are from a high-throughput online serving environment (29,000 examples per second per host), designed to mimic peak production traffic. This setup includes parallel inference streams, dynamic batching, batch queuing, and request de-duplication to accurately reflect online latency reductions.
We use p99 latency (the latency within which 99\% of requests are completed over a 1-minute window) as our evaluation metric.  Real-time, low-latency systems commonly monitor p99 latency to account for worst-case scenarios while excluding outliers.

Figure \ref{fig:copy-time} shows the impact of our optimizations (Section \ref{optimizations}) on the CPU-to-GPU copy time. PinMem denotes the latency incurred from copying to Pinned Memory, and PCIe refers to copy latency over PCIe from PinMem to the GPU. Results are shown for batch sizes of 128 - 512. We report the ablated data for baseline, pinned memory arena only, request de-dup only, and all optimizations combined, making up the 4 bars reported per batch size. Since the pinned memory arena only and all enabled groups directly construct batches on Pinned memory, the PinMem component is zero. With all optimizations, the copy time improves by 85\% for batch size 128, 82\% for 256 and 85\% for 512 compared to the baseline.

Figure \ref{fig:model-runlatency} shows model inference latency improvements.  Model Run latency comprises CPU-to-GPU feature copy, model forward pass, and GPU-to-CPU output copy.  Results are shown for batch sizes 128, 256, and 512, comparing baseline, pinned memory, request deduplication, and the combined approach.
Pinned Memory Arena, when tested independently, has neutral to slightly positive latency reduction results. This is because, on enabling the pinned memory arena exclusively, the setup becomes PCIe bandwidth bound at high throughput. However, Request de-dup only shows consistent reductions across all latencies and batch sizes. The best reduction in Model Run latency is achieved by stacking both optimizations together, resulting in a reduction of 75\%, 75\%, and 81\% for batch sizes 128, 256, 512, respectively, compared to the baseline.

\begin{figure}[h] 
\vspace{-2pt}
  \centering  
  \begin{tikzpicture}  
    \begin{axis}[  
        colormap name=cold,
        point meta min=1.0,
        point meta max=2500.0,
        point meta=explicit,
        ytick={0, 1, 2, 3, 4, 5},  
        xtick={0, 1, 2},  
        xticklabels={p50, p90, p99},  
        yticklabels={Model Forward, Pin Memory, PCIe Copy, Batch Preparation, Queuing Delay, E2E Inference Latency},  
        x tick label style={rotate=0, anchor=east, font=\footnotesize, xshift=1em, yshift=-0.5em},
        y tick label style={font=\footnotesize},
        y axis line style={opacity=0},  
        x axis line style={opacity=0},  
        enlargelimits=false,  
        tick align=outside,  
        tickwidth=0pt,  
        grid style={white},  
        width=0.87\linewidth,  
        height=0.45\linewidth,  
        nodes near coords,
        every node near coord/.append style={
            anchor=east,
            font=\small,
            black,
            xshift=14pt       
        },
        nodes near coords={\pgfmathprintnumber\pgfplotspointmeta$\times$},
    ]  
      
    \addplot [  
        matrix plot,  
        mesh/cols=3, 
    ] table [meta=z] {  
        x  y  z  
        0  0  2.3          
        1  0  2.2           
        2  0  2.2           
        0  1  46            
        1  1  31            
        2  1  37.5          
        0  2  2.5           
        1  2  1.8        
        2  2  1.8          
        0  3  5.7   
        1  3  4.9   
        2  3  4.7   
        0  4  1360       
        1  4  1306        
        2  4  1267   
        0  5  103  
        1  5  338   
        2  5  250  
    };  
    \end{axis}  
  \end{tikzpicture}  
  \caption{Heatmap of Reduction Factors (Baseline vs. Final) }  
  \label{fig:heatmap-final-reducton}  
\end{figure}




Figure \ref{fig:heatmap-final-reducton} shows the improvement ratio of our final optimized setup versus the baseline, across p50, p90, and p99 latency percentiles.  The y-axis breaks down end-to-end RPC latency: model forward, pin memory, and PCIe copy time, batch preparation time (CPU tensor construction) and queuing delay (dynamic batching wait time).  As expected, model forward, pin memory, and PCIe copy latencies improve, in line with the previous results in Figures \ref{fig:copy-time} and \ref{fig:model-runlatency}. Furthermore, request de-dup reduces processed bytes, speeding up batch preparation by a factor of 5x.  Faster mini-batch processing reduces queuing delay by 1300x.  Overall, end-to-end inference latency improves by a factor of 103x, 338x and 250x for p50, p90, p99 respectively.

\section{Conclusions}
\label{sec:conclusion}
This paper introduced TransAct V2, a novel model that integrates real-time and lifelong user sequences to enhance CTR predictions. Leveraging next-action loss, TransAct V2 improves action prediction, boosting engagement and model expressiveness. Its deployment in real-world applications at Pinterest showcases its capability to efficiently manage extensive user histories, tackling latency and storage challenges. Rigorous ablation of serving optimizations demonstrated reduced latency and storage costs, with A/B testing confirming significant improvements in engagement metrics. TransAct V2 advances sequential recommendation, providing a robust framework for deploying complex models at scale.


\bibliographystyle{ACM-Reference-Format}
\balance
\bibliography{main}


\begin{thebibliography}{30}


\ifx \showCODEN    \undefined \def \showCODEN     #1{\unskip}     \fi
\ifx \showDOI      \undefined \def \showDOI       #1{#1}\fi
\ifx \showISBNx    \undefined \def \showISBNx     #1{\unskip}     \fi
\ifx \showISBNxiii \undefined \def \showISBNxiii  #1{\unskip}     \fi
\ifx \showISSN     \undefined \def \showISSN      #1{\unskip}     \fi
\ifx \showLCCN     \undefined \def \showLCCN      #1{\unskip}     \fi
\ifx \shownote     \undefined \def \shownote      #1{#1}          \fi
\ifx \showarticletitle \undefined \def \showarticletitle #1{#1}   \fi
\ifx \showURL      \undefined \def \showURL       {\relax}        \fi
\providecommand\bibfield[2]{#2}
\providecommand\bibinfo[2]{#2}
\providecommand\natexlab[1]{#1}
\providecommand\showeprint[2][]{arXiv:#2}

\bibitem[Borisyuk et~al\mbox{.}(2024)]%
        {lirank}
\bibfield{author}{\bibinfo{person}{Fedor Borisyuk}, \bibinfo{person}{Mingzhou Zhou}, \bibinfo{person}{Qingquan Song}, \bibinfo{person}{Siyu Zhu}, \bibinfo{person}{Birjodh Tiwana}, \bibinfo{person}{Ganesh Parameswaran}, \bibinfo{person}{Siddharth Dangi}, \bibinfo{person}{Lars Hertel}, \bibinfo{person}{Qiang~Charles Xiao}, \bibinfo{person}{Xiaochen Hou}, \bibinfo{person}{Yunbo Ouyang}, \bibinfo{person}{Aman Gupta}, \bibinfo{person}{Sheallika Singh}, \bibinfo{person}{Dan Liu}, \bibinfo{person}{Hailing Cheng}, \bibinfo{person}{Lei Le}, \bibinfo{person}{Jonathan Hung}, \bibinfo{person}{Sathiya Keerthi}, \bibinfo{person}{Ruoyan Wang}, \bibinfo{person}{Fengyu Zhang}, \bibinfo{person}{Mohit Kothari}, \bibinfo{person}{Chen Zhu}, \bibinfo{person}{Daqi Sun}, \bibinfo{person}{Yun Dai}, \bibinfo{person}{Xun Luan}, \bibinfo{person}{Sirou Zhu}, \bibinfo{person}{Zhiwei Wang}, \bibinfo{person}{Neil Daftary}, \bibinfo{person}{Qianqi Shen}, \bibinfo{person}{Chengming Jiang}, \bibinfo{person}{Haichao Wei},
  \bibinfo{person}{Maneesh Varshney}, \bibinfo{person}{Amol Ghoting}, {and} \bibinfo{person}{Souvik Ghosh}.} \bibinfo{year}{2024}\natexlab{}.
\newblock \showarticletitle{LiRank: Industrial Large Scale Ranking Models at LinkedIn}. In \bibinfo{booktitle}{\emph{Proceedings of the 30th ACM SIGKDD Conference on Knowledge Discovery and Data Mining}} (Barcelona, Spain) \emph{(\bibinfo{series}{KDD '24})}. \bibinfo{publisher}{Association for Computing Machinery}, \bibinfo{address}{New York, NY, USA}, \bibinfo{pages}{4804–4815}.
\newblock
\showISBNx{9798400704901}
\urldef\tempurl%
\url{https://doi.org/10.1145/3637528.3671561}
\showDOI{\tempurl}


\bibitem[Chang et~al\mbox{.}(2023)]%
        {TWIN}
\bibfield{author}{\bibinfo{person}{Jianxin Chang}, \bibinfo{person}{Chenbin Zhang}, \bibinfo{person}{Zhiyi Fu}, \bibinfo{person}{Xiaoxue Zang}, \bibinfo{person}{Lin Guan}, \bibinfo{person}{Jing Lu}, \bibinfo{person}{Yiqun Hui}, \bibinfo{person}{Dewei Leng}, \bibinfo{person}{Yanan Niu}, \bibinfo{person}{Yang Song}, {and} \bibinfo{person}{Kun Gai}.} \bibinfo{year}{2023}\natexlab{}.
\newblock \showarticletitle{TWIN: TWo-stage Interest Network for Lifelong User Behavior Modeling in CTR Prediction at Kuaishou}. In \bibinfo{booktitle}{\emph{Proceedings of the 29th ACM SIGKDD Conference on Knowledge Discovery and Data Mining}} (Long Beach, CA, USA) \emph{(\bibinfo{series}{KDD '23})}. \bibinfo{publisher}{Association for Computing Machinery}, \bibinfo{address}{New York, NY, USA}, \bibinfo{pages}{3785–3794}.
\newblock
\showISBNx{9798400701030}
\urldef\tempurl%
\url{https://doi.org/10.1145/3580305.3599922}
\showDOI{\tempurl}


\bibitem[Chen et~al\mbox{.}(2019)]%
        {alibaba_seq_tfmr}
\bibfield{author}{\bibinfo{person}{Qiwei Chen}, \bibinfo{person}{Huan Zhao}, \bibinfo{person}{Wei Li}, \bibinfo{person}{Pipei Huang}, {and} \bibinfo{person}{Wenwu Ou}.} \bibinfo{year}{2019}\natexlab{}.
\newblock \showarticletitle{Behavior Sequence Transformer for E-Commerce Recommendation in Alibaba}. In \bibinfo{booktitle}{\emph{Proceedings of the 1st International Workshop on Deep Learning Practice for High-Dimensional Sparse Data}} (Anchorage, Alaska) \emph{(\bibinfo{series}{DLP-KDD '19})}. \bibinfo{publisher}{Association for Computing Machinery}, \bibinfo{address}{New York, NY, USA}, Article \bibinfo{articleno}{12}, \bibinfo{numpages}{4}~pages.
\newblock
\showISBNx{9781450367837}
\urldef\tempurl%
\url{https://doi.org/10.1145/3326937.3341261}
\showDOI{\tempurl}


\bibitem[Cheng et~al\mbox{.}(2016)]%
        {cheng2016wide}
\bibfield{author}{\bibinfo{person}{Heng-Tze Cheng}, \bibinfo{person}{Levent Koc}, \bibinfo{person}{Jeremiah Harmsen}, \bibinfo{person}{Tal Shaked}, \bibinfo{person}{Tushar Chandra}, \bibinfo{person}{Hrishi Aradhye}, \bibinfo{person}{Glen Anderson}, \bibinfo{person}{Greg Corrado}, \bibinfo{person}{Wei Chai}, \bibinfo{person}{Mustafa Ispir}, {et~al\mbox{.}}} \bibinfo{year}{2016}\natexlab{}.
\newblock \showarticletitle{Wide \& deep learning for recommender systems}. In \bibinfo{booktitle}{\emph{Proceedings of the 1st workshop on deep learning for recommender systems}}. \bibinfo{pages}{7--10}.
\newblock


\bibitem[Dao(2023)]%
        {fav2}
\bibfield{author}{\bibinfo{person}{Tri Dao}.} \bibinfo{year}{2023}\natexlab{}.
\newblock \bibinfo{title}{FlashAttention-2: Faster Attention with Better Parallelism and Work Partitioning}.
\newblock
\newblock
\showeprint[arxiv]{2307.08691}~[cs.LG]
\urldef\tempurl%
\url{https://arxiv.org/abs/2307.08691}
\showURL{%
\tempurl}


\bibitem[Ge et~al\mbox{.}(2020)]%
        {Ge_2020}
\bibfield{author}{\bibinfo{person}{Yingqiang Ge}, \bibinfo{person}{Shuya Zhao}, \bibinfo{person}{Honglu Zhou}, \bibinfo{person}{Changhua Pei}, \bibinfo{person}{Fei Sun}, \bibinfo{person}{Wenwu Ou}, {and} \bibinfo{person}{Yongfeng Zhang}.} \bibinfo{year}{2020}\natexlab{}.
\newblock \showarticletitle{Understanding Echo Chambers in E-commerce Recommender Systems}. In \bibinfo{booktitle}{\emph{Proceedings of the 43rd International ACM SIGIR Conference on Research and Development in Information Retrieval}} \emph{(\bibinfo{series}{SIGIR ’20})}. \bibinfo{publisher}{ACM}, \bibinfo{pages}{2261–2270}.
\newblock
\urldef\tempurl%
\url{https://doi.org/10.1145/3397271.3401431}
\showDOI{\tempurl}


\bibitem[Hertel et~al\mbox{.}(2024)]%
        {hertel2024efficientuserhistorymodeling}
\bibfield{author}{\bibinfo{person}{Lars Hertel}, \bibinfo{person}{Neil Daftary}, \bibinfo{person}{Fedor Borisyuk}, \bibinfo{person}{Aman Gupta}, {and} \bibinfo{person}{Rahul Mazumder}.} \bibinfo{year}{2024}\natexlab{}.
\newblock \bibinfo{title}{Efficient user history modeling with amortized inference for deep learning recommendation models}.
\newblock
\newblock
\showeprint[arxiv]{2412.06924}~[cs.LG]
\urldef\tempurl%
\url{https://arxiv.org/abs/2412.06924}
\showURL{%
\tempurl}


\bibitem[Hsu et~al\mbox{.}(2025)]%
        {liger}
\bibfield{author}{\bibinfo{person}{Pin-Lun Hsu}, \bibinfo{person}{Yun Dai}, \bibinfo{person}{Vignesh Kothapalli}, \bibinfo{person}{Qingquan Song}, \bibinfo{person}{Shao Tang}, \bibinfo{person}{Siyu Zhu}, \bibinfo{person}{Steven Shimizu}, \bibinfo{person}{Shivam Sahni}, \bibinfo{person}{Haowen Ning}, {and} \bibinfo{person}{Yanning Chen}.} \bibinfo{year}{2025}\natexlab{}.
\newblock \bibinfo{title}{Liger Kernel: Efficient Triton Kernels for LLM Training}.
\newblock
\newblock
\showeprint[arxiv]{2410.10989}~[cs.LG]
\urldef\tempurl%
\url{https://arxiv.org/abs/2410.10989}
\showURL{%
\tempurl}


\bibitem[Hsu et~al\mbox{.}(2024)]%
        {twostage}
\bibfield{author}{\bibinfo{person}{Yi-Ping Hsu}, \bibinfo{person}{Po-Wei Wang}, \bibinfo{person}{Chantat Eksombatchai}, {and} \bibinfo{person}{Jiajing Xu}.} \bibinfo{year}{2024}\natexlab{}.
\newblock \showarticletitle{Taming the One-Epoch Phenomenon in Online Recommendation System by Two-stage Contrastive ID Pre-training}. In \bibinfo{booktitle}{\emph{Proceedings of the 18th ACM Conference on Recommender Systems}} (Bari, Italy) \emph{(\bibinfo{series}{RecSys '24})}. \bibinfo{publisher}{Association for Computing Machinery}, \bibinfo{address}{New York, NY, USA}, \bibinfo{pages}{838–840}.
\newblock
\showISBNx{9798400705052}
\urldef\tempurl%
\url{https://doi.org/10.1145/3640457.3688053}
\showDOI{\tempurl}


\bibitem[Kang and McAuley(2018)]%
        {SASRec}
\bibfield{author}{\bibinfo{person}{Wang-Cheng Kang} {and} \bibinfo{person}{Julian McAuley}.} \bibinfo{year}{2018}\natexlab{}.
\newblock \showarticletitle{Self-attentive sequential recommendation}. In \bibinfo{booktitle}{\emph{2018 IEEE international conference on data mining (ICDM)}}. IEEE, \bibinfo{pages}{197--206}.
\newblock


\bibitem[Malreddy et~al\mbox{.}(2024)]%
        {malreddy2024improving}
\bibfield{author}{\bibinfo{person}{Siddarth Malreddy}, \bibinfo{person}{Matthew Lawhon}, \bibinfo{person}{Usha~Amrutha Nookala}, \bibinfo{person}{Aditya Mantha}, {and} \bibinfo{person}{Dhruvil~Deven Badani}.} \bibinfo{year}{2024}\natexlab{}.
\newblock \showarticletitle{Improving feature interactions at Pinterest under industry constraints}.
\newblock \bibinfo{journal}{\emph{arXiv preprint arXiv:2412.01985}} (\bibinfo{year}{2024}).
\newblock


\bibitem[Nguyen et~al\mbox{.}(2021)]%
        {cudagraph}
\bibfield{author}{\bibinfo{person}{Vinh Nguyen}, \bibinfo{person}{Michael Carilli}, \bibinfo{person}{Sukru~Burc Eryilmaz}, \bibinfo{person}{Vartika Singh}, \bibinfo{person}{Michelle Lin}, \bibinfo{person}{Natalia Gimelshein}, \bibinfo{person}{Alban Desmaison}, {and} \bibinfo{person}{Edward Yang}.} \bibinfo{year}{2021}\natexlab{}.
\newblock \showarticletitle{Accelerating PyTorch with CUDA Graphs}.
\newblock  (\bibinfo{year}{2021}).
\newblock
\urldef\tempurl%
\url{https://pytorch.org/blog/accelerating-pytorch-with-cuda-graphs/}
\showURL{%
\tempurl}


\bibitem[Pancha et~al\mbox{.}(2022)]%
        {pinnerformer}
\bibfield{author}{\bibinfo{person}{Nikil Pancha}, \bibinfo{person}{Andrew Zhai}, \bibinfo{person}{Jure Leskovec}, {and} \bibinfo{person}{Charles Rosenberg}.} \bibinfo{year}{2022}\natexlab{}.
\newblock \showarticletitle{PinnerFormer: Sequence Modeling for User Representation at Pinterest}. In \bibinfo{booktitle}{\emph{Proceedings of the 28th ACM SIGKDD Conference on Knowledge Discovery and Data Mining}} (Washington DC, USA) \emph{(\bibinfo{series}{KDD '22})}. \bibinfo{publisher}{Association for Computing Machinery}, \bibinfo{address}{New York, NY, USA}, \bibinfo{pages}{3702–3712}.
\newblock
\showISBNx{9781450393850}
\urldef\tempurl%
\url{https://doi.org/10.1145/3534678.3539156}
\showDOI{\tempurl}


\bibitem[Pi et~al\mbox{.}(2020)]%
        {SIM}
\bibfield{author}{\bibinfo{person}{Qi Pi}, \bibinfo{person}{Xiaoqiang Zhu}, \bibinfo{person}{Guorui Zhou}, \bibinfo{person}{Yujing Zhang}, \bibinfo{person}{Zhe Wang}, \bibinfo{person}{Lejian Ren}, \bibinfo{person}{Ying Fan}, {and} \bibinfo{person}{Kun Gai}.} \bibinfo{year}{2020}\natexlab{}.
\newblock \showarticletitle{Search-based User Interest Modeling with Lifelong Sequential Behavior Data for Click-Through Rate Prediction}.
\newblock \bibinfo{journal}{\emph{Proceedings of the 29th ACM International Conference on Information \& Knowledge Management}} (\bibinfo{year}{2020}).
\newblock
\urldef\tempurl%
\url{https://api.semanticscholar.org/CorpusID:219558850}
\showURL{%
\tempurl}


\bibitem[Pinterest(2024)]%
        {pins}
\bibfield{author}{\bibinfo{person}{Pinterest}.} \bibinfo{year}{2024}\natexlab{}.
\newblock \showarticletitle{Third Quarter 2024 Earnings Conference Call}.
\newblock  (\bibinfo{year}{2024}).
\newblock
\urldef\tempurl%
\url{https://investor.pinterestinc.com/news-and-events/events-and-presentations/event-details/2024/Third-Quarter-2024-Earnings-Conference-Call/default.aspx}
\showURL{%
\tempurl}


\bibitem[Qin et~al\mbox{.}(2020)]%
        {UBR4CTR}
\bibfield{author}{\bibinfo{person}{Jiarui Qin}, \bibinfo{person}{Weinan Zhang}, \bibinfo{person}{Xin Wu}, \bibinfo{person}{Jiarui Jin}, \bibinfo{person}{Yuchen Fang}, {and} \bibinfo{person}{Yong Yu}.} \bibinfo{year}{2020}\natexlab{}.
\newblock \showarticletitle{User Behavior Retrieval for Click-Through Rate Prediction}. In \bibinfo{booktitle}{\emph{Proceedings of the 43rd International ACM SIGIR Conference on Research and Development in Information Retrieval}} \emph{(\bibinfo{series}{SIGIR ’20})}. \bibinfo{publisher}{ACM}, \bibinfo{pages}{2347–2356}.
\newblock
\urldef\tempurl%
\url{https://doi.org/10.1145/3397271.3401440}
\showDOI{\tempurl}


\bibitem[Rabe and Staats(2022)]%
        {memeffattn}
\bibfield{author}{\bibinfo{person}{Markus~N. Rabe} {and} \bibinfo{person}{Charles Staats}.} \bibinfo{year}{2022}\natexlab{}.
\newblock \bibinfo{title}{Self-attention Does Not Need $O(n^2)$ Memory}.
\newblock
\newblock
\showeprint[arxiv]{2112.05682}~[cs.LG]
\urldef\tempurl%
\url{https://arxiv.org/abs/2112.05682}
\showURL{%
\tempurl}


\bibitem[Si et~al\mbox{.}(2024)]%
        {TWINv2}
\bibfield{author}{\bibinfo{person}{Zihua Si}, \bibinfo{person}{Lin Guan}, \bibinfo{person}{Zhongxiang Sun}, \bibinfo{person}{Xiaoxue Zang}, \bibinfo{person}{Jing Lu}, \bibinfo{person}{Yiqun Hui}, \bibinfo{person}{Xingchao Cao}, \bibinfo{person}{Zeyu Yang}, \bibinfo{person}{Yichen Zheng}, \bibinfo{person}{Dewei Leng}, \bibinfo{person}{Kai Zheng}, \bibinfo{person}{Chenbin Zhang}, \bibinfo{person}{Yanan Niu}, \bibinfo{person}{Yang Song}, {and} \bibinfo{person}{Kun Gai}.} \bibinfo{year}{2024}\natexlab{}.
\newblock \showarticletitle{TWIN V2: Scaling Ultra-Long User Behavior Sequence Modeling for Enhanced CTR Prediction at Kuaishou}. In \bibinfo{booktitle}{\emph{Proceedings of the 33rd ACM International Conference on Information and Knowledge Management}} \emph{(\bibinfo{series}{CIKM ’24})}. \bibinfo{publisher}{ACM}, \bibinfo{pages}{4890–4897}.
\newblock
\urldef\tempurl%
\url{https://doi.org/10.1145/3627673.3680030}
\showDOI{\tempurl}


\bibitem[Sun et~al\mbox{.}(2019)]%
        {BERT4Rec}
\bibfield{author}{\bibinfo{person}{Fei Sun}, \bibinfo{person}{Jun Liu}, \bibinfo{person}{Jian Wu}, \bibinfo{person}{Changhua Pei}, \bibinfo{person}{Xiao Lin}, \bibinfo{person}{Wenwu Ou}, {and} \bibinfo{person}{Peng Jiang}.} \bibinfo{year}{2019}\natexlab{}.
\newblock \showarticletitle{BERT4Rec: Sequential Recommendation with Bidirectional Encoder Representations from Transformer}.
\newblock \bibinfo{journal}{\emph{CoRR}}  \bibinfo{volume}{abs/1904.06690} (\bibinfo{year}{2019}).
\newblock
\showeprint[arXiv]{1904.06690}
\urldef\tempurl%
\url{http://arxiv.org/abs/1904.06690}
\showURL{%
\tempurl}


\bibitem[Tillet(2021)]%
        {triton}
\bibfield{author}{\bibinfo{person}{Philippe Tillet}.} \bibinfo{year}{2021}\natexlab{}.
\newblock \showarticletitle{Introducing Triton: Open-source GPU programming for neural networks}.
\newblock  (\bibinfo{year}{2021}).
\newblock
\urldef\tempurl%
\url{https://openai.com/index/triton/}
\showURL{%
\tempurl}


\bibitem[Tillet et~al\mbox{.}(2019)]%
        {triton2}
\bibfield{author}{\bibinfo{person}{Philippe Tillet}, \bibinfo{person}{H.~T. Kung}, {and} \bibinfo{person}{David Cox}.} \bibinfo{year}{2019}\natexlab{}.
\newblock \showarticletitle{Triton: an intermediate language and compiler for tiled neural network computations}. In \bibinfo{booktitle}{\emph{Proceedings of the 3rd ACM SIGPLAN International Workshop on Machine Learning and Programming Languages}} (Phoenix, AZ, USA) \emph{(\bibinfo{series}{MAPL 2019})}. \bibinfo{publisher}{Association for Computing Machinery}, \bibinfo{address}{New York, NY, USA}, \bibinfo{pages}{10–19}.
\newblock
\showISBNx{9781450367196}
\urldef\tempurl%
\url{https://doi.org/10.1145/3315508.3329973}
\showDOI{\tempurl}


\bibitem[Wang et~al\mbox{.}(2024)]%
        {wang2024itemdissimilaritiesdiversifyingintent}
\bibfield{author}{\bibinfo{person}{Yuyan Wang}, \bibinfo{person}{Cheenar Banerjee}, \bibinfo{person}{Samer Chucri}, \bibinfo{person}{Fabio Soldo}, \bibinfo{person}{Sriraj Badam}, \bibinfo{person}{Ed~H. Chi}, {and} \bibinfo{person}{Minmin Chen}.} \bibinfo{year}{2024}\natexlab{}.
\newblock \bibinfo{title}{Beyond Item Dissimilarities: Diversifying by Intent in Recommender Systems}.
\newblock
\newblock
\showeprint[arxiv]{2405.12327}~[cs.IR]
\urldef\tempurl%
\url{https://arxiv.org/abs/2405.12327}
\showURL{%
\tempurl}


\bibitem[Wijmans et~al\mbox{.}(2024)]%
        {cce}
\bibfield{author}{\bibinfo{person}{Erik Wijmans}, \bibinfo{person}{Brody Huval}, \bibinfo{person}{Alexander Hertzberg}, \bibinfo{person}{Vladlen Koltun}, {and} \bibinfo{person}{Philipp Krähenbühl}.} \bibinfo{year}{2024}\natexlab{}.
\newblock \bibinfo{title}{Cut Your Losses in Large-Vocabulary Language Models}.
\newblock
\newblock
\showeprint[arxiv]{2411.09009}~[cs.LG]
\urldef\tempurl%
\url{https://arxiv.org/abs/2411.09009}
\showURL{%
\tempurl}


\bibitem[Wilhelm et~al\mbox{.}(2018)]%
        {ddp}
\bibfield{author}{\bibinfo{person}{Mark Wilhelm}, \bibinfo{person}{Ajith Ramanathan}, \bibinfo{person}{Alexander Bonomo}, \bibinfo{person}{Sagar Jain}, \bibinfo{person}{Ed~H. Chi}, {and} \bibinfo{person}{Jennifer Gillenwater}.} \bibinfo{year}{2018}\natexlab{}.
\newblock \showarticletitle{Practical Diversified Recommendations on YouTube with Determinantal Point Processes}. In \bibinfo{booktitle}{\emph{Proceedings of the 27th ACM International Conference on Information and Knowledge Management}} (Torino, Italy) \emph{(\bibinfo{series}{CIKM '18})}. \bibinfo{publisher}{Association for Computing Machinery}, \bibinfo{address}{New York, NY, USA}, \bibinfo{pages}{2165–2173}.
\newblock
\showISBNx{9781450360142}
\urldef\tempurl%
\url{https://doi.org/10.1145/3269206.3272018}
\showDOI{\tempurl}


\bibitem[Xia et~al\mbox{.}(2023)]%
        {transact}
\bibfield{author}{\bibinfo{person}{Xue Xia}, \bibinfo{person}{Pong Eksombatchai}, \bibinfo{person}{Nikil Pancha}, \bibinfo{person}{Dhruvil~Deven Badani}, \bibinfo{person}{Po-Wei Wang}, \bibinfo{person}{Neng Gu}, \bibinfo{person}{Saurabh~Vishwas Joshi}, \bibinfo{person}{Nazanin Farahpour}, \bibinfo{person}{Zhiyuan Zhang}, {and} \bibinfo{person}{Andrew Zhai}.} \bibinfo{year}{2023}\natexlab{}.
\newblock \showarticletitle{TransAct: Transformer-based Realtime User Action Model for Recommendation at Pinterest}. In \bibinfo{booktitle}{\emph{Proceedings of the 29th ACM SIGKDD Conference on Knowledge Discovery and Data Mining}} (Long Beach, CA, USA) \emph{(\bibinfo{series}{KDD '23})}. \bibinfo{publisher}{Association for Computing Machinery}, \bibinfo{address}{New York, NY, USA}, \bibinfo{pages}{5249–5259}.
\newblock
\showISBNx{9798400701030}
\urldef\tempurl%
\url{https://doi.org/10.1145/3580305.3599918}
\showDOI{\tempurl}


\bibitem[Xu et~al\mbox{.}(2022)]%
        {rethink}
\bibfield{author}{\bibinfo{person}{Jiajing Xu}, \bibinfo{person}{Andrew Zhai}, {and} \bibinfo{person}{Charles Rosenberg}.} \bibinfo{year}{2022}\natexlab{}.
\newblock \showarticletitle{Rethinking Personalized Ranking at Pinterest: An End-to-End Approach}. In \bibinfo{booktitle}{\emph{Proceedings of the 16th ACM Conference on Recommender Systems}} (Seattle, WA, USA) \emph{(\bibinfo{series}{RecSys '22})}. \bibinfo{publisher}{Association for Computing Machinery}, \bibinfo{address}{New York, NY, USA}, \bibinfo{pages}{502–505}.
\newblock
\showISBNx{9781450392785}
\urldef\tempurl%
\url{https://doi.org/10.1145/3523227.3547394}
\showDOI{\tempurl}


\bibitem[Xu et~al\mbox{.}(2024)]%
        {jaggedattn}
\bibfield{author}{\bibinfo{person}{Rengan Xu}, \bibinfo{person}{Junjie Yang}, \bibinfo{person}{Yifan Xu}, \bibinfo{person}{Hong Li}, \bibinfo{person}{Xing Liu}, \bibinfo{person}{Devashish Shankar}, \bibinfo{person}{Haoci Zhang}, \bibinfo{person}{Meng Liu}, \bibinfo{person}{Boyang Li}, \bibinfo{person}{Yuxi Hu}, \bibinfo{person}{Mingwei Tang}, \bibinfo{person}{Zehua Zhang}, \bibinfo{person}{Tunhou Zhang}, \bibinfo{person}{Dai Li}, \bibinfo{person}{Sijia Chen}, \bibinfo{person}{Gian-Paolo Musumeci}, \bibinfo{person}{Jiaqi Zhai}, \bibinfo{person}{Bill Zhu}, \bibinfo{person}{Hong Yan}, {and} \bibinfo{person}{Srihari Reddy}.} \bibinfo{year}{2024}\natexlab{}.
\newblock \showarticletitle{Enhancing Performance and Scalability of Large-Scale Recommendation Systems with Jagged Flash Attention}. In \bibinfo{booktitle}{\emph{18th ACM Conference on Recommender Systems}} \emph{(\bibinfo{series}{RecSys ’24})}. \bibinfo{publisher}{ACM}, \bibinfo{pages}{778–780}.
\newblock
\urldef\tempurl%
\url{https://doi.org/10.1145/3640457.3688040}
\showDOI{\tempurl}


\bibitem[Yan et~al\mbox{.}(2024)]%
        {trinity}
\bibfield{author}{\bibinfo{person}{Jing Yan}, \bibinfo{person}{Liu Jiang}, \bibinfo{person}{Jianfei Cui}, \bibinfo{person}{Zhichen Zhao}, \bibinfo{person}{Xingyan Bin}, \bibinfo{person}{Feng Zhang}, {and} \bibinfo{person}{Zuotao Liu}.} \bibinfo{year}{2024}\natexlab{}.
\newblock \bibinfo{title}{Trinity: Syncretizing Multi-/Long-tail/Long-term Interests All in One}.
\newblock
\newblock
\showeprint[arxiv]{2402.02842}~[cs.IR]
\urldef\tempurl%
\url{https://arxiv.org/abs/2402.02842}
\showURL{%
\tempurl}


\bibitem[Ying et~al\mbox{.}(2018)]%
        {PinSage}
\bibfield{author}{\bibinfo{person}{Rex Ying}, \bibinfo{person}{Ruining He}, \bibinfo{person}{Kaifeng Chen}, \bibinfo{person}{Pong Eksombatchai}, \bibinfo{person}{William~L. Hamilton}, {and} \bibinfo{person}{Jure Leskovec}.} \bibinfo{year}{2018}\natexlab{}.
\newblock \showarticletitle{Graph Convolutional Neural Networks for Web-Scale Recommender Systems}. In \bibinfo{booktitle}{\emph{Proceedings of the 24th ACM SIGKDD International Conference on Knowledge Discovery and Data Mining}} (London, United Kingdom) \emph{(\bibinfo{series}{KDD '18})}. \bibinfo{publisher}{Association for Computing Machinery}, \bibinfo{address}{New York, NY, USA}, \bibinfo{pages}{974–983}.
\newblock
\showISBNx{9781450355520}
\urldef\tempurl%
\url{https://doi.org/10.1145/3219819.3219890}
\showDOI{\tempurl}


\bibitem[Zhai et~al\mbox{.}(2024)]%
        {hstu}
\bibfield{author}{\bibinfo{person}{Jiaqi Zhai}, \bibinfo{person}{Lucy Liao}, \bibinfo{person}{Xing Liu}, \bibinfo{person}{Yueming Wang}, \bibinfo{person}{Rui Li}, \bibinfo{person}{Xuan Cao}, \bibinfo{person}{Leon Gao}, \bibinfo{person}{Zhaojie Gong}, \bibinfo{person}{Fangda Gu}, \bibinfo{person}{Michael He}, \bibinfo{person}{Yinghai Lu}, {and} \bibinfo{person}{Yu Shi}.} \bibinfo{year}{2024}\natexlab{}.
\newblock \bibinfo{title}{Actions Speak Louder than Words: Trillion-Parameter Sequential Transducers for Generative Recommendations}.
\newblock
\newblock
\showeprint[arxiv]{2402.17152}~[cs.LG]
\urldef\tempurl%
\url{https://arxiv.org/abs/2402.17152}
\showURL{%
\tempurl}


\end{thebibliography}

\appendix
\section{Additional Ablation Studies On NAL}
\subsection{NAL Loss Weight}
We tuned loss weight $w_{NAL}$ in Eq~\ref{equation: loss}. As shown in Table~\ref{table:nal_lw}, adjusting the NAL loss weight is necessary for maintaining a balance between the Next Action Loss task and the multi-head prediction task. This equilibrium is essential because the next action prediction serves as an auxiliary task, whereas multi-head prediction is the primary objective of the ranking model. If $w_{NAL}$ is too large, the model's performance deteriorates on multi-head prediction due to the disproportionate influence of NAL. Conversely, if $w_{NAL}$ is too small, the contribution of NAL becomes negligible. Based on our ablation study, we determined that setting $w_{NAL}=0.01$ is optimal for our model. However, it is worth noting that the ideal weight may vary across different models and datasets, as it needs to accommodate the diverse magnitudes of loss values encountered in various applications.

\begin{table}
\caption{Offline evaluation of NAL loss weight tuning (impression-based negatives). ($^*$ statistically insignificant) }
  \resizebox{.45\textwidth}{!}{%
  \begin{tabular}{ccccc}
    \toprule
    NAL loss weight & HIT@3/repin & HIT@3/hide \\
    \midrule
   0.0001 & +0.58\% & -2.09\% \\
   0.001 & +0.86\% & \textbf{-2.6\%} \\
   0.01 & \textbf{+1.10\%} & -2.39\% \\
   0.1 & +0.54\% & -2.39\% \\
  \bottomrule
\end{tabular}%
}
\label{table:nal_lw}
\end{table}

\subsection{NAL Loss Type}
Table~\ref{table:nal_loss_type} shows that the sampled softmax loss outperforms the cross-entropy loss. 
Cross-entropy loss is typically the most popular choice for classification tasks, such as predicting positive or negative actions in our setup. However, sampled softmax loss offers greater flexibility by allowing fine-tuning of the positive-to-negative sample size ratio, as illustrated in Equation~\ref{equation: SSM}. This additional freedom contributes to consistently enhanced performance.

As demonstrated in Table~\ref{table:nal_loss_type}, the sampled softmax loss leads to superior outcomes compared to cross-entropy loss. This advantage underscores the efficacy of adjusting sampling ratios to achieve improved classification results.

\begin{table}
\caption{Offline evaluation of NAL loss types. ($^*$ statistically insignificant)}
  \resizebox{.45\textwidth}{!}{%
  \begin{tabular}{ccccc}
    \toprule
    NAL loss type & HIT@3/repin & HIT@3/hide \\
    \midrule
   cross entropy & \textbf{-1.31\%} & \textbf{0.27\%}$^*$ \\
   sampled softmax & \textbf{0.18\%} & \textbf{-1.84\%} \\
  \bottomrule
\end{tabular}%
}
\label{table:nal_loss_type}
\end{table}

\section{Additional Ablation Studies on SKUT}
\label{appendix_skut}
The comparison presented in Table~\ref{tab:latency-memory-decrease-bs} underscores SKUT's superiority over PyTorch's memory-efficient implementation across various batch sizes, with a fixed sequence length of 192. Although FlashAttention-2 is not directly applicable to our work due to its lack of support for custom masking, we conducted a performance comparison with SKUT in the absence of masking. Our SKUT implementation demonstrates a substantial advantage, achieving 66.4\% lower latency and 5.5\% reduced memory usage compared to a FlashAttention-2 transformer without attention masks.
\begin{table}[h]
    \centering
    \caption{Latency and Memory Reduction by Batch Size (SKUT vs. PyTorch~\cite{memeffattn})}



\begin{tabular}{|c|c|c|}  
    \hline  
    \textbf{Batch Size} & \textbf{Latency Decrease} & \textbf{Memory Decrease} \\  
    \hline  
    16   & -60.36\% & -1.34\% \\  
    \hline  
    32   & -65.96\% & -3.20\% \\  
    \hline  
    64   & -76.79\% & -4.45\% \\  
    \hline  
    128  & -82.79\% & -8.80\% \\  
    \hline  
    \textbf{256}  & \textbf{-85.09\%} & \textbf{-13.24\%} \\  
    \hline  
    512  & -86.69\% & -23.01\% \\  
    \hline  
    1024 & -87.28\% & -31.47\% \\  
    \hline  
    2048 & -87.79\% & -38.55\% \\  
    \hline  
\end{tabular}  

    \label{tab:latency-memory-decrease-bs}
\end{table}

\end{document}